\documentclass[a4paper,article,10pt]{memoir}
\usepackage[english]{babel}
\usepackage{CCLAuthor}
\usepackage[reqno]{amsmath}
\usepackage{amssymb}
\usepackage{amsthm}  
\theoremstyle{plain}

\theoremstyle{definition}

\numberwithin{theorem}{chapter}
\usepackage{amsmath,amsfonts,amssymb,epsfig,psfrag,subfigure}
\usepackage{amsmath,amsfonts,amssymb,,epsfig}

\renewcommand{\le}{\leqslant}
\renewcommand{\ge}{\geqslant}

\newcommand{\be}{\begin{equation}}
\newcommand{\en}{\end{equation}}

\newcommand{\ii}{\textrm{i}}
\newcommand{\ee}{\textrm{e}}
\renewcommand{\vec}[1]{\boldsymbol{#1}}
  
\usepackage[round,authoryear]{natbib} 
\begin{document}
%
%
%
\title{Interface Waves \\in Pre-Stressed Incompressible Solids}
%
%
\author{%
    Michel Destrade
    \\ \smallskip\small
    Institut Jean Le Rond d'Alembert, CNRS/Universit\'e Pierre et Marie Curie, Paris, France}
%
%
    \maketitle
%
%
%
    \vskip-2\baselineskip\noindent \begin{abstract}
    We study incremental wave propagation for what is seemingly the simplest
    boundary value problem, namely that constitued by the plane interface of 
    a semi-infinite solid.
    With a view to model loaded elastomers and soft tissues, we focus on incompressible 
    solids, subjected to large homogeneous static deformations.
    The resulting strain-induced anisotropy complicates matters for the 
    incremental boundary value problem, but we transpose and take advantage 
    of powerful techniques and results from the linear anisotropic elastodynamics theory. 
    In particular we cover several situations where fully explicit secular equations can 
    be derived, including Rayleigh and Stoneley waves in principal directions, 
    and Rayleigh waves polarized in a principal plane or propagating in any
    direction in a principal plane. 
    We also discuss the merits of polynomial secular equations with respect to 
    more robust, but less transparent, exact secular equations.
    
    \end{abstract}
    
    \CCLsection{Introduction}
    The term ``acousto-elastic effect'' describes the interplay between the static 
    deformation of an elastic solid and the motion of an elastic wave.
    If both the deformation and the motion are of infinitesimal amplitude,
    then all the governing equations are linearized, see for instance the Chapter 
    by Norris for examples and applications or the experimental results of 
    \citet{PaSF84}. 
    If both the deformation and the motion are of finite amplitude, then the 
    resulting governing equations are highly nonlinear,
    and their resolution is the subject of much research, 
    see the Chapter by Fu for the weakly nonlinear theory and the 
    Chapter by Saccomandi for the fully nonlinear theory. 
    
    In between those two situations lies the theory of ``small-on-large'',
    also known as the theory of ``incremental'' motions, where the wave is 
    an infinitesimal perturbation superimposed onto the large static homogeneous deformation 
    of a generic hyperelastic solid.
    There, the homogeneous character of the static deformation and the linear character
    of the incremental equations of motion ensure that the calculations are valid for 
    any strain energy density (to be specified later for applications, if necessary).
    The next Section of this Chapter briefly recalls the governing equations of incremental
    motions (see the Chapter by Ogden for their derivation).
    
    It turns out that many similarities can be drawn between the  
    equations of incremental motions and those of linear anisotropic elasticity,
    with the main difference that in the latter case, the anisotropy is set once and 
    for all for a given crystal whereas in the former case, it is strain-induced 
    and susceptible to great variations from one configuration to another.
    Using the similarities, we may transpose the so-called Stroh formulation and
    exploit its many results;
    on the other hand, when focussing on the differences, we may highlight the 
    influences of the pre-stress and of the choice of a strain-energy density on 
    the propagation of waves. 
    In this Chapter, attention is restricted to waves at the interface of pre-deformed,
    semi-infinite solids, in contact either with vacuum (Rayleigh waves) or with another 
    solid (Stoneley waves).
    With a view to model elastomers and biological soft tissues, the solids are considered to 
    be incompressible 
    (mathematically, this internal constraint lightens somewhat the expressions but 
    does not prove essential to the resolution).
    
    Several situations are treated:
    principal wave propagation in Section 3, principal polarization in Section 4,
    and principal plane propagation in Section 5. 
    The emphasis is on deriving explicit secular equations in polynomial form,
    using some simple ``fundamental equations'' derived at the end of Section 2. 
    Of course, as the setting gets more and more involved, so does the search for 
    a polynomial secular equation;
    eventually its degree becomes too high for comfort and other techniques are required.
    The concluding section (Section 6) discusses the pros and cons of such equations, 
    as opposed to exact, non-explicit, secular equations, free of spurious roots.

    \CCLsection{Basic equations}
    
    \CCLsubsection{Finite deformation}

Consider an isotropic, incompressible, hyperelastic solid at rest, 
characterized by a mass density $\rho$ and a  strain energy function $W$.
Then subject it to a large, static, homogeneous deformation (``the pre-strain'')
carrying the particle at $\vec{X}$ in the undeformed configuration
to the position $\vec{x}$ in the deformed configuration. 

\begin{figure} [htbp]
\centering 
\epsfig{figure=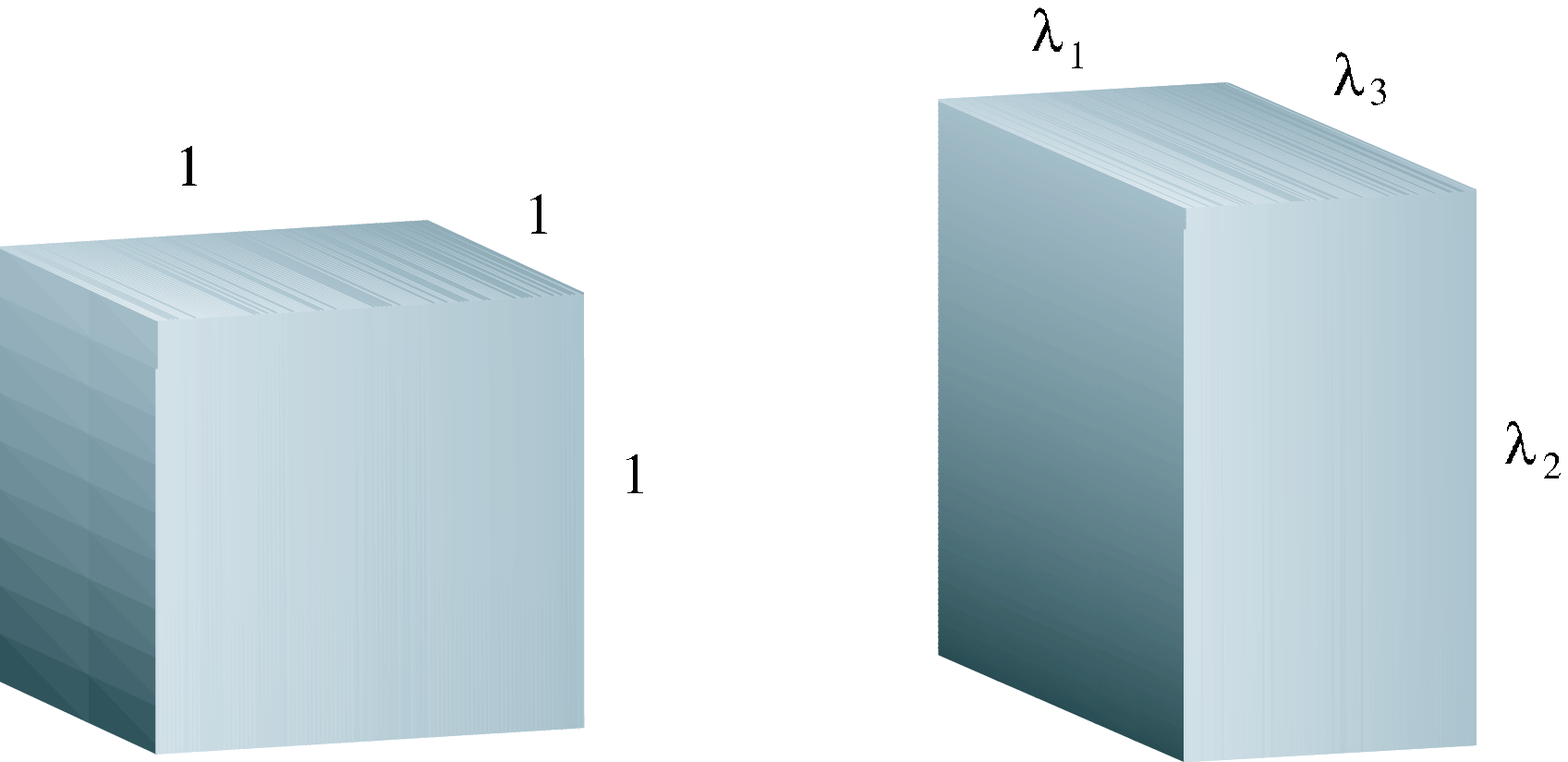,  width=.9\textwidth}
 \caption{
 Finite homogeneous deformation of a unit cube.\label{fig_1}}
\end{figure}

Call $\vec{F} = \partial \vec{x} / \partial \vec{X}$ the corresponding \emph{constant}
deformation gradient and $\vec{B} = \vec{F F}^t$ the associated left Cauchy-Green 
strain tensor. 
This tensor being symmetric, the directions of its eigenvectors are
orthogonal;
they are called the principal axes of pre-strain 
or in short, the \emph{principal axes}. 
Also, the eigenvalues of $\vec{B}$ are positive, 
$\lambda_1^2$, $\lambda_2^2$, $\lambda_3^2$, say, and 
$\lambda_1$, $\lambda_2$, $\lambda_3$ are called the 
\emph{principal stretches}.
Figure \ref{fig_1} shows how a unit cube with edges aligned with the principal 
axes, is transformed by the pre-strain.
Note that because the solid is \emph{incompressible}, its volume is 
preserved through any deformation so that here,
\be \label{incompressibility}
\lambda_1 \lambda_2 \lambda_3 =1.
\en

The first two \emph{principal invariants} of strain are defined as
\be
I_1 = \text{tr }\vec{B}, \quad 
I_2 = [(\text{tr }\vec{B})^2 - \text{tr }(\vec{B}^2)]/2. 
\en
In the Cartesian coordinate system aligned with the principal 
axes, $\vec{B}$ is diagonal. 
Calling $\vec{e}_1$, $\vec{e}_2$, $\vec{e}_3$, the unit vectors 
in the $x_1$, $x_2$, $x_3$ directions, respectively, we have 
\be 
\vec{B} = \lambda_1^2 \vec{e}_1 \otimes \vec{e}_1
  + \lambda_2^2  \vec{e}_2 \otimes \vec{e}_2
   + \lambda_3^2 \vec{e}_3 \otimes \vec{e}_3,
\en
and the computation of $I_1$, $I_2$ there gives
\be \label{I-lambda}
I_1 = \lambda_1^2 + \lambda_2^2 + \lambda_3^2, \quad
I_2 = \lambda_1^2 \lambda_2^2 + \lambda_2^{2} \lambda_3^2 + \lambda_3^{2} \lambda_1^2.
\en

For an \emph{isotropic} solid, $W$ may be given as a function of the 
invariants: $W = W(I_1, I_2)$ or equivalently, as a symmetric function of the 
principal stretches: $W = W(\lambda_1, \lambda_2, \lambda_3)$, 
according to what is most convenient for the analysis or 
according to 
how $W$ has been determined experimentally. 
With the first choice, the \emph{constant} Cauchy stress 
(``the pre-stress'') necessary to maintain the solid in its 
state of finite homogeneous deformation is 
\be \label{prestress_I}
\vec{\sigma} = -p \vec{I} 
  + 2(\partial W/\partial I_1 + I_1 \partial W/\partial I_2) \vec{B} 
   - 2(\partial W / \partial I_2) \vec{B}^{2},
\en
where $p$ is a Lagrange multiplier due to the constraint of 
incompressibility (a yet arbitrary constant scalar to be determined 
from initial and boundary conditions.)
With the second choice, the non-zero components of 
$\vec{\sigma}$ relative to the principal axes are written as 
\be \label{pre-stress}
\sigma_i = -p + \lambda_i \partial W / \partial \lambda_i,
\qquad 
i = 1, 2, 3 \text{ (no sum)}.
\en
The proof for the equivalence between \eqref{prestress_I} and \eqref{pre-stress} 
relies on the connections \eqref{I-lambda}.
    
    \CCLsubsection{Incremental equations}
    \label{2_2}
    
Consider a half-space filled with an incompressible 
hyperelastic solid subject to a large homogeneous 
deformation.
We take the Cartesian coordinate system 
($\hat{x}_1$, $\hat{x}_2$, $\hat{x}_3$) to be oriented 
so that the boundary is at $\hat{x}_2 = 0$, 
and we study the propagation in the $\hat{x}_1$ direction of an infinitesimal 
\emph{interface wave} in the solid. 

This wave is \emph{inhomogeneous} as it progresses in an harmonic 
manner in a direction lying in the interface, while its amplitude 
decays with distance from the boundary. 

We call $\vec{u}$ the \emph{mechanical displacement} associated with the 
wave, and $\dot{p}$ the increment in the Lagrange multiplier $p$ due to 
incompressibility. 
The \emph{incremental nominal stress} tensor $\vec{s}$ has components 
\be \label{s}
s_{j i} = \mathcal{A}_{0jilk} u_{k,l} + p u_{j,i} - \dot{p} \delta_{i j},
\en
where the comma denotes partial differentiation with respect to the  
coordinates $\hat{x}_1$, $\hat{x}_2$, $\hat{x}_3$. 
Here, $\vec{\mathcal{A}_0}$ is the fourth-order tensor of 
\emph{instantaneous elastic moduli}, with components 
\be
\mathcal{A}_{0jilk} = 
 F_{j \alpha} F_{l \beta} \dfrac{\partial^2 W}{F_{i \alpha} F_{k \beta}}
   = \mathcal{A}_{0lkji}.
\en
Note that due to the symmetry above, $\vec{\mathcal{A}_0}$ has in general 45 
independent components.  
In the principal axes coordinate system ($x_1 x_2 x_3$) however, 
there are only 15 independent non-zero components;
they are \citep{Ogde01}:
\begin{align} \label{A0_principal}
& \mathcal{A}_{0iijj} =
\lambda_i \lambda_j W_{ij}, &&
\nonumber \\
& \mathcal{A}_{0ijij} =
  (\lambda_i W_i-\lambda_j W_j)\lambda_i^2/(\lambda_i^2-\lambda_j^2),
&& i \ne j, \quad \lambda_i \ne \lambda_j,
\nonumber \\
& \mathcal{A}_{0ijij} =
  (\mathcal{A}_{0iiii} - \mathcal{A}_{0iijj} + \lambda_i W_i)/2,
&& i \ne j, \quad \lambda_i = \lambda_j,
\nonumber \\
& \mathcal{A}_{0ijji}= \mathcal{A}_{0jiij}=
   \mathcal{A}_{0ijij} - \lambda_i W_i, && i \ne j, 
\end{align}
(no sums on repeated indexes here),
where $W_j = \partial W / \partial \lambda_j$ and 
$W_{i j} = \partial^2 W / (\partial \lambda_i \partial \lambda_j)$.

Finally, the \emph{governing equations} are the 
incremental equations of motion and the incremental 
constraint of incompressibility; they read
\be \label{governing}
s_{j i,j} = \rho \partial^2 u_i / \partial t^2, \qquad 
 u_{j,j} = 0,
\en
respectively.

Now everything is in place to solve an interface wave problem.
We take $\vec{u}$ and $\dot{p}$ in the form 
\be \label{U_P}
\{ \vec{u}, \dot{p} \} = \{ \vec{U}(k \hat{x}_2), \ii k P(k \hat{x}_2) \}
  \ee^{\ii k (\vec{n \cdot \hat{x}} - v t)},
\en
where $k$ is the wave number, $\vec{U}$ and $P$ are functions of the variable 
$k \hat{x}_2$ only, $\vec{n}$ is the unit vector in the direction of propagation, 
and $v$ is the speed. 
Clearly by \eqref{s}, $\vec{s}$ has a similar form, say
\be \label{S}
\vec{s} = \ii k \vec{S}(k \hat{x}_2) \ee^{\ii k (\vec{n \cdot \hat{x}} - v t)},
\en
where $\vec{S}$ is a function of $k \hat{x}_2$ only.

After substitution, it turns out that the 
incremental governing equations \eqref{governing} can be cast as 
the following first-order differential system \citep{Chad97},
\be \label{motion}
\vec{\xi}' = \ii \vec{N} \vec{\xi}, \qquad 
\text{where} \quad 
\vec{\xi} = [\vec{U}, \vec{t}]^t,
\en
the prime denotes differentiation with respect to the variable 
$k \hat{x}_2$, and $\vec{t}$ are the \emph{tractions} acting on planes 
parallel to the boundary, with components $t_j = S_{2j}$.
Here the matrix $\vec{N}$ has the following block structure
\begin{equation} \label{N}
\vec{N}  =  \begin{bmatrix}
                    \vec{N}_1 & \vec{N}_2 \\
   \vec{N}_3 +\rho v^2 \vec{I}   & \vec{N}_1^t
                             \end{bmatrix},
\end{equation}
where $\vec{N}_1$, $\vec{N}_2 = \vec{N}_2^t$,
and $\vec{N}_3 = \vec{N}_3^t$ are square matrices.
This is the so-called \emph{Stroh formulation}. 
In effect, many of the results established thanks 
to the \citet{Stro62} formalism in linear anisotropic elasticity can 
formally be carried over 
to the context of incremental dynamics in nonlinear elasticity, as shown by 
\cite{ChJa79a}, \cite{Chad97}, and \citet{Fu05a, Fu05b}.

    \CCLsubsection{Resolution}

The solution to the first-order differential system \eqref{motion} 
is an exponential function in $k \hat{x}_2$, 
\be \label{expo}
\vec{\xi}(k \hat{x}_2) =  \ee^{\ii k q \hat{x}_2} \vec{\zeta},
\en
where $\vec{\zeta}$ is  a constant vector and $q$ is a scalar. 
Then the following eigenvalue problem emerges:
$\vec{N} \vec{\zeta} = q \vec{\zeta}$. 
Its resolution is in two steps.

First, find the eigenvalues by solving the \emph{propagation condition},
\be \label{propagation_condition}
\text{det }(\vec{N} - q \vec{I}) = 0,
\en
for $q$, 
and keep those $q_j$'s which satisfy the \emph{decay condition}.
For instance, when the solid fills up the $\hat{x}_2 \ge 0$ half-space, 
the decay condition is 
\be \label{decay}
\Im (q) > 0,
\en
ensuring that the solution \eqref{expo} 
is localized near the interface and vanishes away from it.
The \emph{penetration depth} of the interface wave is clearly 
related to the magnitude of $\Im (q)$:
the smaller this quantity is, the deeper the wave penetrates into the 
solid.

The propagation condition \eqref{propagation_condition} 
is a polynomial in $q$ with real coefficients and 
it has only complex roots \citep{Fu05b}, which come 
therefore in pairs of complex conjugate quantities. 
Hence, half of all the roots to the propagation condition 
qualify as satisfying the decay condition.
Let $\vec{\zeta}^j$ be the eigenvector corresponding to the 
qualifying root $q_j$.

Now proceed to the second step, which is to construct 
the general localized solution to the equations of motion, 
as 
\be \label{general_solution}
\vec{\xi}(k \hat{x}_2) = 
   \textstyle{\sum} \gamma_j  \ee^{\ii k q_j \hat{x}_2} \vec{\zeta}^j,
\en
for some arbitrary constants $\gamma_j$. 
Then compute this vector at the interface $\hat{x}_2 = 0$ and 
apply the \emph{boundary conditions}. 
The vector $\vec{\xi}(0)$ is often 
decomposed as follows,
\be
\vec{\xi}(0) = 
 \begin{bmatrix}
  \vec{U}(0) \\ \vec{t}(0) \end{bmatrix}
   = \textstyle{\sum} \gamma_j \vec{\zeta}^j
    =  \begin{bmatrix}
          \vec{A} \\ \vec{B} \end{bmatrix}
        \vec{\gamma},
\en
where $\vec{A}$ and $\vec{B}$ are square matrices and $\vec{\gamma}$ 
is the vector with components $\gamma_j$.
For instance, the archetype of interface waves is the \citet{Rayl85}
surface wave, which propagates at the interface between a solid 
half-space and the vacuum, leaving the boundary free of tractions. 
Mathematically, the corresponding boundary condition is that 
$\vec{t}(0) = \vec{B \gamma} = \vec{0}$, leading to 
\be \label{secularB}
\text{det } \vec{B} = 0.
\en
This (complex) form of the \emph{secular equation} is however not the optimal 
form, and it might lead to unsatisfactory answers to the 
questions of existence and uniqueness of the wave (see 
\cite{Barn00} for an historical account of this point).
From the Stroh formalism, and its application to the present context, 
we learn that it is much more efficient to work with the 
\emph{surface impedance matrix} \citep{InTo69}
than with the matrices $\vec{A}$ and $\vec{B}$; 
this matrix $\vec{M}$ is defined by 
\be
\vec{M} = - \ii \vec{B A}^{-1}.
\en
It is Hermitian \citep{BaLo85, Fu05b} and so 
$\text{det }\vec{M} = -\ii (\text{det }\vec{B}) / (\text{det }\vec{A})$
is a real quantity and the \emph{secular equation for Rayleigh surface 
waves}, written in the form
\be \label{secularM}
\text{det } \vec{M} = 0,
\en
is a \emph{real} equation, in contrast to \eqref{secularB}. 
Moreover, if there is a root to this equation in the subsonic regime (where $v$ is less 
than the speed of any bulk wave), then it is unique; 
also, the existence of a root is equivalent to the existence of a surface wave.

Similar results also exist for other types of interface waves as seen in the 
course of this Chapter. 
For instance, the boundary conditions for \citet{Ston24} interface waves
are that displacements and tractions are continuous across the boundary 
between two rigidly bonded semi-infinite solids. 
Then $\vec{\xi}(0) = \vec{\xi}^*(0)$ where the asterisk refers to quantities 
for the solid in $\hat{x}_2 \le 0$. 
Equivalently, $\vec{A \gamma} = \vec{A}^* \vec{\gamma}^*$, 
$\vec{B \gamma} = \vec{B}^* \vec{\gamma}^*$, from which comes
$[\vec{BA}^{-1} \vec{A}^* -  \vec{B}^*] \vec{\gamma}^* =  \vec{0}$,
leading to 
\be \label{stoneley}
\text{det} (\vec{M + M}^*) = 0,
\en 
the optimal form of the \emph{secular equation for Stoneley interface waves}.
Here $\vec{M}^*$ is the surface impedance matrix for the 
solid in the $\hat{x}_2 \le 0$ half-space, defined as
\be
\vec{M}^* =  \ii \vec{B}^* (\vec{A}^*)^{-1}.
\en

    \CCLsubsection{Explicit secular equations}

The derivation of a secular equation, preferably in the 
optimal form involving the surface impedance matrix, is no 
sinecure in general. 
The problematic step lies in the resolution of the propagation 
condition \eqref{propagation_condition}.

For principal wave propagation 
($\hat{x}_1$, $\hat{x}_2$, $\hat{x}_3$ are aligned with the principal axes),
the propagation condition factorizes into the product of a term linear in 
$q^2$ and a term quadratic in $q^2$. 
Here we can compute the roots explicitly, keep the qualifying ones
(see \eqref{decay}), and solve the boundary value problem in its 
entirety. 
Many problems falling in this category have been solved over the years, 
and some are presented in Section \ref{section_3}. 

For a non-principal wave with propagation direction and attenuation 
direction both in a principal plane 
(the saggital plane ($\hat{x}_1 \hat{x}_2$) is a principal plane but 
$\hat{x}_1$ is not a principal axis), the propagation condition factorizes 
 into the product of a term linear in 
$q^2$ and a term quartic in $q$.
We treat this case in Section \ref{section_4}.
Although it is possible to write down formally the qualifying 
roots of the quartic \citep{Fu05a, DeFu06, FuBr06}, 
the formulas involved are cumbersome to interpret. 

For a wave propagating in a principal plane but not in a principal direction 
($\hat{x}_2$ is aligned with a principal axis but neither $\hat{x}_1$ nor $\hat{x}_3$ are 
aligned with principal axes), 
the propagation condition is a cubic in $q^2$.
We treat this case in Section \ref{section_5}.
Now it is a daunting task to find analytical expressions for the roots 
$q$ satisfying the decay condition \eqref{decay}.

Finally, for wave propagation in any other case, the propagation 
condition is a sextic in $q$, unsolvable analytically according to 
Galois theory.

These observations suggest that, except in the case of principal waves, 
numerical procedures are required in order to make progress. 
It is indeed the case that sophisticated tools and efficient numerical recipes 
have been developed by \citet{BaLo85}, 
\citet{FuMi02}, and several others, with most satisfying results. 
However it is also the case that some interface wave problems 
can be solved analytically, up to the derivation of the secular 
equation in \emph{explicit polynomial form}. 
The first steps in that direction were taken by \citet{Curr79},
and his advances were later refined by \citet{TaCu81}
and \citet{Tazi89}, revisited by \citet{Mozh95} and by \citet{Ting04}, and extended by 
\citet{Dest03}.

The equations that turn out to  be fundamental in the derivation 
of explicit polynomial secular equations are
\be \label{fundamental}
\vec{\xi}(0) \cdot \vec{\hat{I}}\vec{N}^n \overline{\vec{\xi}}(0) = 0,
\en
where $\vec{\hat{I}} = \begin{bmatrix} \vec{0} & \vec{I} \\ \vec{I} & \vec{0} \end{bmatrix}$
and $n$ is an integer. 
Their derivation is most simple.
First, it can be shown by induction \citep{Ting04} that $\vec{N}^n$ has a block structure 
similar to that of $\vec{N}$, that is 
\begin{equation} \label{Nn}
\vec{N}^n     =  \begin{bmatrix}
                    \vec{N}^{(n)}_1 & \vec{N}^{(n)}_2 \\
                    \vec{K}^{(n)}   & \vec{N}^{(n)t}_1
              \end{bmatrix},
\end{equation}
with $\vec{K}^{(n)} = \vec{K}^{(n)t}$, 
$\vec{N}_2^{(n)} = \vec{N}^{(n)t}_2$.
It then follows that 
\be
\vec{\hat{I} N}^n =  \begin{bmatrix}
                       \vec{K}^{(n)}   & \vec{N}^{(n)t}_1 \\
                       \vec{N}^{(n)}_1 & \vec{N}^{(n)}_2
                     \end{bmatrix}
\end{equation}
is \emph{symmetric} for all $n$.
Now take the scalar product of both sides of the 
governing equation \eqref{motion} by 
$\vec{\hat{I} N}^n \overline{\vec{\xi}}$ to get
\be 
\vec{\xi}' \cdot \vec{\hat{I}}\vec{N}^n \overline{\vec{\xi}} = 
\ii \vec{\xi} \cdot \vec{\hat{I}}\vec{N}^{n+1} \overline{\vec{\xi}};
\en
finally add its complex conjugate to this equality to end up with
\be 
\vec{\xi}' \cdot \vec{\hat{I}}\vec{N}^n \overline{\vec{\xi}} + 
 \vec{\xi} \cdot \vec{\hat{I}}\vec{N}^{n} \overline{\vec{\xi}}' = 0,
\en
and, by integration between the interface (at $\hat{x}_2 = 0$) and infinity 
(where $\vec{U}$ and $\vec{t}$, and thus $\vec{\xi}$, vanish), 
arrive at \eqref{fundamental}. 

For instance, the boundary condition for Rayleigh surface waves is  
that there are no incremental tractions at the interface;
thus $\vec{\xi}(0) = [\vec{U}(0), \vec{0}]^t$, and the fundamental 
equations \eqref{fundamental} reduce to 
\begin{equation} \label{fundamental_rayleigh}
  \overline{\vec{U}}(0) \cdot \vec{K}^{(n)} \vec{U}(0)=0.
\end{equation}

    \CCLsection{Principal waves}
    \label{section_3}

\begin{figure}  [htbp]
\centering 
\epsfig{figure=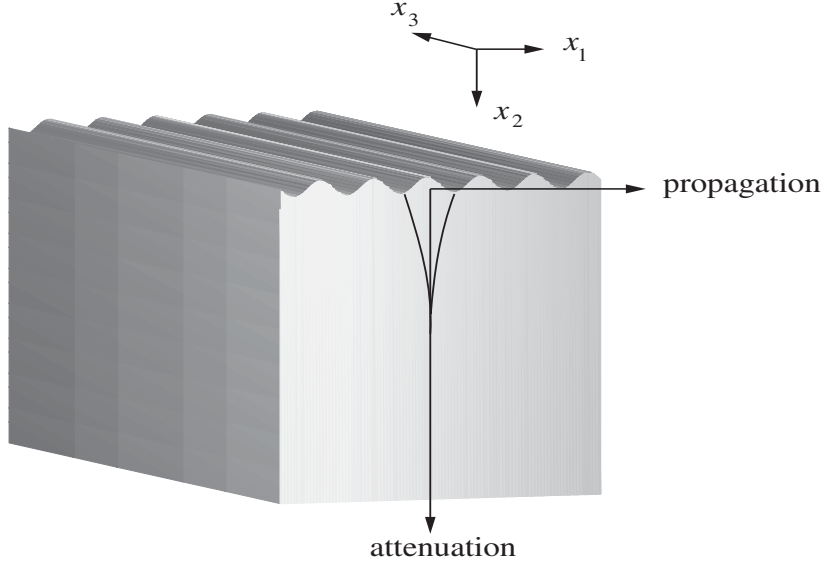, width=.8\textwidth, height=.4\textheight}
 \caption{
 Incremental wave propagation localized 
 near the surface of a semi-infinite deformed solid. 
 The analysis does not give the amplitude nor the 
 wavelength. \label{figure_2}}
\end{figure}

Here we take ($\hat{x}_1, \hat{x}_2, \hat{x}_3$) to coincide
with the principal axes ($x_1, x_2, x_3$).
The pre-deformation is thus
\be \label{triaxial}
\hat{x}_1 =  \lambda_1 X_1, \quad
\hat{x}_2 =  \lambda_2 X_2, \quad
\hat{x}_3 =  \lambda_3 X_3.
\en
Figure \ref{figure_2} summarizes the situation with respect to the  
waves' characteristics near the interface. 
Bear in mind that the wave analysis is linear and gives no indication 
about the amplitude; 
moreover, a half-space has no characteristic length 
so that the secular equation is non-dispersive and the 
wavelength remains undetermined. 
 
    \CCLsubsection{Governing equations}
    \label{section_3_1}

For principal waves, 
the fields \eqref{U_P} and \eqref{S} are independent of $\hat{x}_3 = x_3$,
because $\hat{x}_2 = x_2$ and $\vec{n \cdot \hat{x}} = \hat{x}_1 = x_1$.
Also, recall from \eqref{A0_principal} that 
the non-zero components of $\vec{\mathcal{A}_0}$ in the ($x_1, x_2, x_3$) 
coordinate system of principal axes are 
\begin{align} \label{A-principal}
& \mathcal{A}_{01111} = \lambda_1^2 W_{11}, 
\quad
\mathcal{A}_{01122} = \lambda_1 \lambda_2 W_{12}, 
\quad
\mathcal{A}_{02222} = \lambda_2^2 W_{22}, 
\notag \\
& \lambda_1^{-2} \mathcal{A}_{01212} =
\lambda_2^{-2} \mathcal{A}_{02121} = 
\dfrac{ \lambda_1 W_1 - \lambda_2 W_2}{\lambda_1^2 - \lambda_2^2}, 
\quad
\mathcal{A}_{01221} = 
\dfrac{ \lambda_2 W_1 - \lambda_1 W_2}{\lambda_1^2 - \lambda_2^2} \lambda_1 \lambda_2,
\end{align}
and also $\mathcal{A}_{03333}$, $\mathcal{A}_{01133}$, $\mathcal{A}_{02233}$, 
$\mathcal{A}_{01313}$, $\mathcal{A}_{02323}$,
$\mathcal{A}_{03131}$, $\mathcal{A}_{03232}$,
$\mathcal{A}_{01331}$, and $\mathcal{A}_{02332}$, whose expressions 
are not needed in this Section.      

From these observations follows that the third equation of motion 
\eqref{governing}$_3$ reduces to 
\be
- S_{13} + \ii S'_{23} = - \rho v^2 U_3,
\en
where by \eqref{s},
\be
\ii S_{13} = \ii \mathcal{A}_{01313} U_3, \qquad 
\ii S_{23} = \mathcal{A}_{02323} U_3'.
\en
Hence the movement along the $x_3$ principal axis is governed by 
an equation which depends only on $U_3$. 
For this equation, governing what is termed the \emph{anti-plane}
motion, we take the trivial solution: $U_3 = 0$, and we focus on 
the \emph{in-plane} motion. 
According to \eqref{governing}$_{1,2,4}$, it is governed by
\be \label{in-plane}
- S_{11} + \ii S'_{21} = - \rho v^2 U_1, \qquad
- S_{12} + \ii S'_{22} = - \rho v^2 U_2, \qquad 
\ii U_1 + U'_2 = 0,
\en 
where by  \eqref{s},
\begin{align}
& \ii S_{11} = \ii (\mathcal{A}_{01111} + p) U_1 + \mathcal{A}_{01122} U'_2 -P,
\notag \\
& \ii S_{21} = \mathcal{A}_{02121} U'_1 + \ii (\mathcal{A}_{01221} + p) U_2,
\notag \\
& \ii S_{12} = (\mathcal{A}_{01221} + p) U'_1 + \ii \mathcal{A}_{01212} U_2,
\notag \\
& \ii S_{22} = \ii \mathcal{A}_{01122} U_1 + (\mathcal{A}_{02222} + p)U'_2 -P.
\end{align}
We eliminate $p$ in favour of the pre-stress: by \eqref{pre-stress} 
at $j=2$, we have $p = \lambda_2 W_2 - \sigma_2$ and so by \eqref{A-principal},
\be
\mathcal{A}_{01221} + p = \mathcal{A}_{02121} - \sigma_2.
\en
It then follows from the second equation above that
\be
U'_1 = \ii \left[ - \dfrac{\mathcal{A}_{02121} - \sigma_2}{\mathcal{A}_{02121}} U_2
        + \dfrac{1}{\mathcal{A}_{02121}} S_{21} \right],
\en
and this constitutes the first line of the first-order system \eqref{motion}.
The second line comes from the incremental incompressibility constraint 
\eqref{in-plane}$_3$ as 
\be
U'_2 = \ii \left[ - U_1 \right].
\en
Proceeding similarly for $S'_{21}$, $S'_{22}$, 
we find eventually that the governing equations are indeed in the 
form \eqref{motion}, where $\vec{\xi} = [U_1, U_2, S_{21}, S_{22}]^t$ 
and $-\vec{N}_1$,  $\vec{N}_2$, and $-\vec{N}_3$ are given by
\begin{equation}  \label{N1N2N3_principal}
 \begin{bmatrix}
       0 &  \dfrac{\gamma_{21}-\sigma_2}{\gamma_{21}} \\
       1 & 0 
 \end{bmatrix},
\quad
 \begin{bmatrix}
       \dfrac{1}{\gamma_{21}} & 0 \\
           0 & 0 
  \end{bmatrix},
\quad
 \begin{bmatrix}
     2 (\beta_{12} + \gamma_{21} - \sigma_2)   &     0 \\
      0    &     \gamma_{12}
      - \dfrac{(\gamma_{21}-\sigma_2)^2}{\gamma_{21}}
 \end{bmatrix},
\end{equation}
respectively, where  we used the following 
short-hand notations (no sums),
\begin{align} 
& \gamma_{i j} =
  \mathcal{A}_{0ijij} = \lambda_i^2 \lambda_j^{-2} \gamma_{j i},
\nonumber \\
& 2 \beta_{i j} =  
  \mathcal{A}_{0iiii} + \mathcal{A}_{0jjjj} 
    - 2\mathcal{A}_{0iijj} - 2\mathcal{A}_{0ijji}  
                = 2 \beta_{j i}.
\end{align}
or equivalently,
\begin{align} \label{gammaBeta}
& \gamma_{i j} =
  (\lambda_i W_i-\lambda_j W_j)\lambda_i^2/(\lambda_i^2-\lambda_j^2)
  = \lambda_i^2 \lambda_j^{-2} \gamma_{j i},
\nonumber \\
& 2 \beta_{i j} = \lambda_i^2 W_{ii}
   - 2 \lambda_i \lambda_j W_{i j} + \lambda_j^2 W_{j j}
 +  2(\lambda_i W_j-\lambda_j W_i)
         \lambda_i \lambda_j/(\lambda_i^2-\lambda_j^2)
                = 2 \beta_{j i}.
\end{align}

    \CCLsubsection{Resolution}
    \label{section_3_2}

The \emph{propagation condition} \eqref{propagation_condition}
reduces to a quadratic in $q^2$,
\be \label{biquadratic}
\gamma_{21} q^4 + (2 \beta_{12} - \rho v^2) q^2 + \gamma_{12} - \rho v^2 = 0.
\en
Notice how $\sigma_2$, though present in $\vec{N}$, does not appear 
explicitly in this equation.

Calling $q_1^2$, $q_2^2$, the roots of the quadratic, we have
\be
q_1^2 q_2^2 = \dfrac{\gamma_{12} - \rho v^2}{\gamma_{21}}, \quad
q_1^2 + q_2^2 = -\dfrac{2\beta_{12} - \rho v^2}{\gamma_{21}}.
\en

The roots $q_1$, $q_2$ of the biquadratic satisfying the decay condition 
\eqref{decay} are in one of the two following forms;
either: $q_1 = \ii \beta_1$, $q_2 = \ii \beta_2$, where $\beta_1 > 0$, 
$\beta_2 > 0$, or: $q_1 = \alpha + \ii \beta$, 
 $q_2 = -\alpha + \ii \beta$, where $\beta > 0$. 
Whatever the case, $q_1^2 q_2^2 > 0$, $q_1 q_2 <0$, 
and $q_1 + q_2$ is a purely imaginary quantity. 
From the first inequality we deduce that \citep{DoOg90}
\be
\eta = \sqrt{\dfrac{\gamma_{12} - \rho v^2}{\gamma_{21}}}
\en
is a real quantity. 
From the second, and using the definitions of $\eta$ and $\gamma_{i j}$, 
we find
\be \label{q1q2_q1+q2}
q_1 q_2 = - \eta, \quad 
(q_1 + q_2)^2 = \dfrac{\gamma_{12} - 2\beta_{12}}{\gamma_{21}} - 2\eta - \eta^2
 = \lambda_1^2 \lambda_2^{-2} - 2\dfrac{\beta_{12}}{\gamma_{21}} - 2\eta - \eta^2.
\en

We compute the eigenvectors $\vec{\zeta}^1$ and $\vec{\zeta}^2$ of $\vec{N}$ 
corresponding to $q_1$ and $q_2$ as any column of the matrix adjoint to 
$\vec{N} - q_1 \vec{I}$ and to $\vec{N} - q_2 \vec{I}$, respectively.
Choosing the third column, we find
\be
\vec{\zeta}^1 = \begin{bmatrix}\vec{a}^1 \\ \vec{b}^1\end{bmatrix}, \quad
\vec{\zeta}^2 = \begin{bmatrix}\vec{a}^2 \\ \vec{b}^2\end{bmatrix}, 
\en
where
\be
\vec{a}^j = \left[- \dfrac{q_j^2}{\gamma_{21}}, \dfrac{q_j}{\gamma_{21}}\right]^t, 
\quad
\vec{b}^j = \left[- q_j(q_j^2 - 1 + \overline{\sigma}_2), 
  q_j^2(1 - \overline{\sigma}_2) - \eta^2\right]^t, 
\en
and $\overline{\sigma}_2 = \sigma_2 / \gamma_{21}$ is a 
non-dimensional measure of the pre-stress.

We can now construct the $\vec{A} = [\vec{a}^1 | \vec{a}^2]$ and 
$\vec{B} = [\vec{b}^1| \vec{b}^2]$ matrices, and the surface impedance 
matrix $\vec{M} = - \ii \vec{B A}^{-1}$.
It turns out to be 
\be
\vec{M} = - \ii \gamma_{21} 
 \begin{bmatrix}
  q_1 + q_2 & 1 - \overline{\sigma}_2 - \eta \\
  -(1 - \overline{\sigma}_2 - \eta) & (q_1 + q_2)\eta
  \end{bmatrix},
\en
which is indeed Hermitian because $q_1 + q_2$ is a purely imaginary quantity
and $\eta$ is real.

For \emph{Rayleigh surface waves}, the secular equation is \eqref{secularM},
or here, using \eqref{q1q2_q1+q2},
\be \label{secular_principal}
\eta^3 + \eta^2 + 
  (2 - \lambda_1^2 \lambda_2^{-2} 
    + 2\dfrac{\beta_{12}}{\gamma_{21}} 
     - 2 \overline{\sigma}_2)\eta
       - (1-\overline{\sigma}_2)^2 = 0.
\en
\citet{DoOg90} established this form of the secular equation 
for principal surface waves in pre-stressed incompressible solids, following 
other works by \citet{HaRi61}, \citet{Flav63}, 
\citet{Will73, Will74}, \citet{ChJa79a}, 
\citet[review with an extensive bibliography]{Guz02}, 
and many others.
It is of course consistent with Lord Rayleigh's own analysis 
of surface waves in linear isotropic incompressible solids. 
To check this, let the solid be un-stressed ($\sigma_i = 0$) 
and un-deformed ($\lambda_i = 1$);
then $\eta$ reduces to $\sqrt{1 - \rho v^2 / \mu_0}$, 
where $\mu_0$ is the infinitesimal shear modulus;
also, $\beta_{12} = \gamma_{21}$ and $\eta$ is the 
real root of $\eta^3 + \eta^2 + 3\eta - 1 =0$, that is 
$\eta \simeq 0.2956$ giving $\rho v^2 / \mu_0 \simeq 0.9126$, 
as found by \citet{Rayl85}.

For \emph{Stoneley interface waves}, the secular equation is 
 \eqref{stoneley} where 
\be
\vec{M + M}^* = - \ii  
 \begin{bmatrix}
  \gamma_{21}(q_1 + q_2) - \gamma_{21}^*(q_1^* + q_2^*) 
    & \gamma_{21}(1 - \eta) - \gamma_{21}^*(1 - \eta^*) \\
  -\gamma_{21}(1 - \eta) + \gamma_{21}^*(1 - \eta^*) 
    & \gamma_{21}(q_1 + q_2)\eta - \gamma^*_{21}(q_1^* + q_2^*)\eta^*
  \end{bmatrix}.
\en
This equation was studied in great detail by \citet{DoOg91} and by \citet{Chad95}. 
It is consistent with the analysis of \citet{Ston24} 
of interface waves in linear isotropic incompressible solids.
A remarkable feature of this secular equation for principal 
Stoneley interface waves in deformed incompressible solids -- first noted by 
\citet{ChJa79b} -- is that the pre-stress $\sigma_2$ does not 
appear explicitly in it, in contrast to the equation for surface waves
\eqref{secular_principal}.  
This quantity, which is continuous across the interface ($\sigma_2 = \sigma_2^*$),
disappears in the addition of the two surface impedance matrices. 
Of course it still plays an implicit role, in determining the pre-strain.

\citet{DoOg90, DoOg91}, \citet{Chad95}, and \citet[review]{Guz02} have covered 
almost every aspect  of principal interface wave propagation and 
more information can be found in their respective articles. 
In the next Subsection we rapidly work out two examples of surface 
waves.

    \CCLsubsection{Examples}
    \label{section_3_3}

First we present an example taken from the literature on elastomers,
where the Mooney-Rivlin strain energy function is often encountered. 
It is given by
\be \label{mooney}
W = \mathcal{D}_1 (\lambda_1^2 + \lambda_2^2 + \lambda_3^2 -3)/2
  + \mathcal{D}_2 (\lambda_1^2 \lambda_2^2
                     + \lambda_2^2 \lambda_3^2 + \lambda_3^2\lambda_1^2 -3)/2,
\en
where $\mathcal{D}_1$ and $\mathcal{D}_2$ are positive constants  
with the dimensions of a stiffness.
The Mooney-Rivlin material enjoys special properties 
with respect to  wave propagation 
(the neo-Hookean material, 
which corresponds to the special case $\mathcal{D}_2 = 0$, enjoys even 
more special properties as is seen in Section \ref{section_5_3_1}).
For instance, once subjected to a large homogeneous pre-strain, 
it permits the propagation of bulk waves \emph{in every direction}; 
these waves can be infinitesimal, but also of 
arbitrary \emph{finite amplitude} \citep{BoHa92};
they can be homogeneous plane waves but also \emph{inhomogeneous} 
plane waves \citep{Dest00, Dest02}. 
The quantities \eqref{gammaBeta} are also quite special; 
they are
\begin{equation}
\gamma_{ij}
  = (\mathcal{D}_1 + \mathcal{D}_2 \lambda_k^2)\lambda_i^2,
\quad
      2\beta_{ij} =
 (\mathcal{D}_1 + \mathcal{D}_2 \lambda_k^2)(\lambda_i^2+\lambda_j^2),
\end{equation}
where $k \ne i,j$, and thus they satisfy 
\begin{equation} \label{relations_mooney}
     2\beta_{ij} = \gamma_{ij} + \gamma_{ji}.
\end{equation}
These relationships mean that the biquadratic  \eqref{biquadratic}
factorizes to 
\be
(q^2 + 1)(q^2 + \eta^2) = 0,
\en
and that the secular equation \eqref{secular_principal} reduces to 
\be \label{secular_MR}
\eta^3 + \eta^2 + 
  (3 - 2 \overline{\sigma}_2)\eta
       - (1-\overline{\sigma}_2)^2 = 0.
\en
Hence one qualifying root is $q_1 = \ii$, whatever the values 
of the material constants $\mathcal{D}_1$ and $\mathcal{D}_2$. 
The other root is $q_2 = \ii \eta$.
When there is no pre-stress normal to the boundary ($\overline{\sigma}_2 = 0$), 
then $\eta$ is the 
real root of $\eta^3 + \eta^2 + 3\eta - 1 =0$, that is 
$\eta \simeq 0.2956$ giving 
\be \label{principal_mr}
\rho v^2 = \gamma_{12} - \gamma_{21}\eta^2
 = (\mathcal{D}_1 + \mathcal{D}_2 \lambda_3^2)
     (\lambda_1^2 - 0.0874 \lambda_2^2),
\en
a result first established by \citet{Flav63}.
Here $q_1 = \ii$ and $q_2 \simeq 0.295 \ii$ so that the 
penetration depth of the surface wave is fixed and is 
\emph{completely independent of the pre-strain and of the material parameters} 
$\mathcal{D}_1$ and $\mathcal{D}_2$.
We say that the penetration depth is universal relative to the class of 
Mooney-Rivlin materials.

\begin{figure}  [htbp]
\centering 
\epsfig{figure=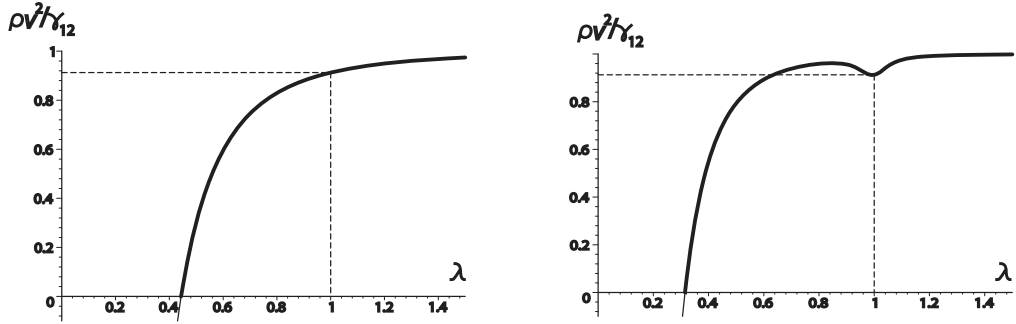, width=.99\textwidth}
 \caption{
 Variations of the scaled squared surface wave speed with the stretch in 
 uniaxial pre-stress  (a) for any Mooney-Rivlin material and (b) 
 for the solid with strain energy density  \eqref{Vorp} where 
  $\mathcal{C}_1 = 0.175$ MPa, 
$\mathcal{C}_2 = 1.9$ MPa.\label{figure_3}}
\end{figure}

The second example is taken from the biomechanics literature.
From a series of uniaxial tests on human aortic aneurysms, 
\citet{RaVo00} deduced that the following strain energy density 
gave a satisfying fit with the data plots,
\begin{equation} \label{Vorp}
W = \mathcal{C}_1 (\lambda_1^2 + \lambda_2^2 +\lambda_3^2 -3)
    + \mathcal{C}_2 (\lambda_1^2 + \lambda_2^2 +\lambda_3^2 -3)^2,
\end{equation}
where, typically, $\mathcal{C}_1 = 0.175$ MPa, 
$\mathcal{C}_2 = 1.9$ MPa 
(\citet{KaHY97} use the same expression to model the response
of passive myocardium.) 
A uniaxial pre-stress is $\sigma_1 \ne 0$, $\sigma_2 = \sigma_3 = 0$,
leading through \eqref{pre-stress} to the following equi-biaxial pre-strain,
\begin{equation} \label{equi-biaxial}
 \lambda_1 = \lambda, \quad \lambda_2 = \lambda^{-1/2}, 
 \quad \lambda_3 = \lambda^{-1/2},
\end{equation}
where $\lambda$ is calculated from 
\be
\sigma_1 = 2 (\lambda^2 - \lambda^{-1})
  [\mathcal{C}_1 + 2 \mathcal{C}_2(2 - 3 \lambda^{-1} + \lambda^{-3})].
\en
Finally, using the following expressions for the relevant moduli,
\begin{align}
& \gamma_{21} = 
 2\mathcal{C}_1 \lambda^{-1} + 4 \mathcal{C}_2(2\lambda^{-1} - 3 \lambda^{-2} + \lambda^{-4}),
 \qquad
 \gamma_{12} = \lambda^3 \gamma_{21},
 \notag \\
&\beta_{21} = 
 \mathcal{C}_1 (\lambda^2 + \lambda^{-1})
   + 2 \mathcal{C}_2(4\lambda^{2} - 3 \lambda - \lambda^{-1} - 3 \lambda^{-2} + 3\lambda^{-4}),
\end{align}
it is a simple matter to solve the secular equation 
\eqref{secular_principal} numerically and 
plot the variations of the squared wave speed, scaled with respect to 
the squared bulk wave speed $\gamma_{12}/\rho$, with the pre-stretch $\lambda$. 
Figure \ref{figure_3}$b$ displays these variations;
for comparison purposes, Figure \ref{figure_3}$a$ shows the variations of the 
scaled squared wave speed in the case of a Mooney-Rivlin material in 
uniaxial stress; 
in that later case the graph is independent of the 
material parameters $\mathcal{D}_1$ and $\mathcal{D}_2$ because 
by \eqref{principal_mr}, $\rho v^2 / \gamma_{21} = 1 - 0.0874 \lambda^{-3}$.
The dashed lines indicate the speed of Lord Rayleigh's squared speed in the 
isotropic (no pre-strain) case where $\lambda = 1$, $\rho v^2 / \gamma_{12} \simeq 0.9126$. 
Notice how different the responses of two solids are in that neighbourhood. 
Note also that for high compressive stretches, the squared speeds eventually falls 
off to zero;
this happens at $\lambda \simeq 0.0874^{1/3} \simeq 0.444$ for all Mooney-Rivlin 
materials, as shown by \cite{Biot63},
and at  $\lambda \simeq 0.315$ for the soft biological tissue model above. 
Beyond that \emph{critical compression stretch}, $v^2 < 0$,
leading to a purely imaginary $v$, an amplitude which then grows 
exponentially with time according to  \eqref{U_P}, and a breakdown 
of the linearized analysis. 
The search for critical compression stretches is an extremely active 
area of research, clearly linked to the geometric stability analysis of solids. 

    \CCLsection{Waves polarized in a principal plane}
    \label{section_4}

In this Section we study the case where $\hat{x}_3$ is aligned 
with the principal axis $x_3$ but neither $\hat{x}_1$ (propagation direction)
nor  $\hat{x}_2$ (attenuation direction) are aligned with principal 
axes. 
The components $\hat{\mathcal{A}}_{0jilk}$ in the 
($\hat{x}_1 \hat{x}_2 \hat{x}_3$) coordinate system of the 
instantaneous elastic moduli tensor $\vec{\mathcal{A}_0}$ 
are related to the components $\mathcal{A}_{0jilk}$, given by 
\eqref{A-principal},  in the 
($x_1 x_2 x_3$) coordinate system of principal axes
 through the tensor transformations
\begin{equation} \label{omega}
\hat{\mathcal{A}}_{0jilk} = 
 \Omega_{j p} \Omega_{i q} \Omega_{l r} \Omega_{k s} \mathcal{A}_{0pqrs}, \quad
 \text{where} \quad 
 \vec{\Omega} =
   \begin{bmatrix}
     \cos \Theta & -\sin \Theta & 0 \\
    \sin \Theta & \cos \Theta & 0 \\
          0      &      0      & 1
    \end{bmatrix},
\end{equation}
and $\Theta$ is the angle between $x_1$ and $\hat{x}_1$.

In particular we find that the non-zero components in the forms 
 $\hat{\mathcal{A}}_{013lk}$ and $\hat{\mathcal{A}}_{023lk}$
are  $\hat{\mathcal{A}}_{01313}$, $\hat{\mathcal{A}}_{01323}$,
 $\hat{\mathcal{A}}_{01331}$, $\hat{\mathcal{A}}_{02323}$,  
 $\hat{\mathcal{A}}_{02313}$, and $\hat{\mathcal{A}}_{02332}$.
The mechanical fields \eqref{U_P} and \eqref{S} are independent of 
$\hat{x}_3 = x_3$ because $\vec{n \cdot \hat{x}} = \hat{x}_1$ here.
As a consequence, the third equation of motion 
\eqref{governing}$_3$ reduces to 
\be
- S_{13} + \ii S'_{23} = - \rho v^2 U_3,
\en
where by \eqref{s},
\be
\ii S_{13} = \ii \hat{\mathcal{A}}_{01313} U_3 + \hat{\mathcal{A}}_{01323} U_3', 
\quad 
\ii S_{23} =  \ii \hat{\mathcal{A}}_{02313} U_3 + \hat{\mathcal{A}}_{02323} U_3'.
\en
Hence the movement along the $x_3$ principal axis is governed by 
an equation which depends only on $U_3$,  
and for this anti-plane motion we take the trivial solution: $U_3 = 0$.

The equations governing the in-plane motion have been derived in the 
case of a general plane pre-strain by \citet{Fu05a} and solved for surface waves 
in the case of a pre-strain consisting in a triaxial stretch 
followed by a simple shear by \citet{DeOg05}. 
Instead of treating these cases again, we revisit the 
case relative to one of the most important pre-strain fitting 
into the present context, that of \emph{finite simple shear},
presented originally by \citet{CoOg95} for surface waves.

\begin{figure}  [htbp]
\centering 
 \epsfig{figure=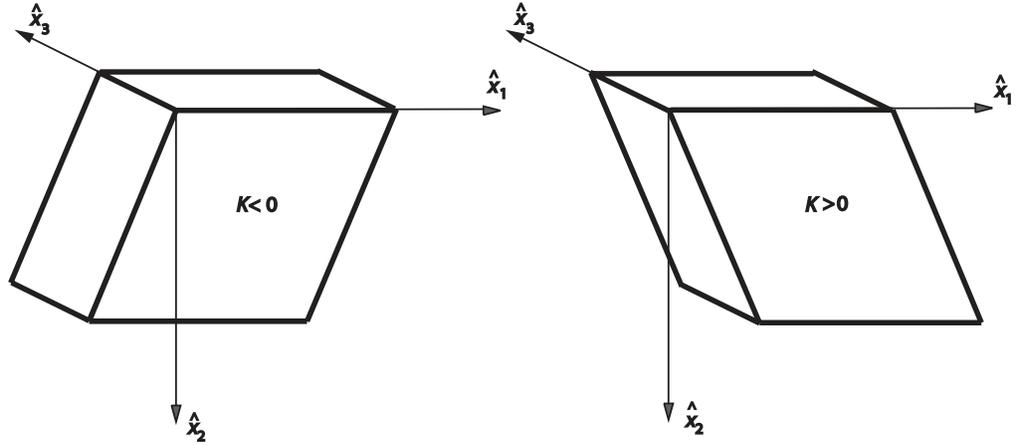, width=.99\textwidth}
 \caption{
 Finite simple shear of amount $K$ of a block near the interface. 
 The $\hat{x}_1$ direction is the \emph{direction of shear};  
 the ($\hat{x}_1 \hat{x}_2$) plane is the \emph{plane of shear}; 
 the ($\hat{x}_1 \hat{x}_3$) plane is the \emph{glide plane}. \label{figure_4}}
\end{figure}

Figure \ref{figure_4} sketches what happens to a unit cube when a solid is 
subject to the simple shear of amount $K$,
\be \label{shear}
\hat{x}_1 =  X_1 + K X_2, \quad
\hat{x}_2 =   X_2, \quad
\hat{x}_3 =   X_3.
\en
Here the principal axes are $x_3 = X_3$  and $x_1$, $x_2$ which 
make an angle $\psi$ with $X_1$ and with $X_2$, respectively.
That angle, and the corresponding principal stretches are 
(e.g. \cite{Chad76}),
\begin{equation}
  \psi =  (1/2) \tan^{-1} (2 / K), \quad
  \lambda_{1,2} = \sqrt{1 + K^2/4} \pm K/2, \quad
\lambda_3 = 1.
\end{equation} 
These relations highlight a major difference between this 
homogeneous pre-strain and the triaxial pre-stretch \eqref{triaxial}:
here the orientation of the principal axes with respect to 
the plane interface changes as the magnitude 
of the pre-strain changes.

    \CCLsubsection{Governing equations for simple shear pre-strain}

The deformation \eqref{shear} is an example of \emph{plane strain}. 
As we focus on two-partial incremental waves in this Section, 
we may take advantage of formulas established by \citet{MeOg02}
in a similar context.
The components of the deformation gradient tensor $\vec{F}$
and of the left Cauchy-Green strain tensor $\vec{B}$ 
for 2D pre-strain and 2D incremental motions, 
in the ($\hat{x}_1, \hat{x}_2$) coordinate system 
(aligned with the ($X_1, X_2$) system), are
\be \label{B_shear}
\vec{F} = \begin{bmatrix} 1 & K \\
                          0 & 1 
          \end{bmatrix}, 
\qquad
\vec{B} = \begin{bmatrix} 1 + K^2& K \\
                          K & 1 
          \end{bmatrix}.
\en

For plane strain, $\lambda_3 = 1$, so that in incompressible solids
$\lambda_2 = \lambda_1^{-1}$ by \eqref{incompressibility}. 
It follows by \eqref{I-lambda} that $I_1 = I_2$.
Accordingly, we define the single-variable function $\widehat{W}(I_1)$ by the identity
\be \label{What}
\widehat{W}(I_1) = W(I_1, I_1).
\en
Then the 2D version of the constitutive equation 
\eqref{prestress_I} is 
\be \label{prestress_plane}
\vec{\sigma} = - \hat{p} \vec{I} 
  + 2 \widehat{W}_1 \vec{B},
\en
where $\hat{p}$ is the Lagrange multiplier due to the incompressibility constraint.
Also, the components of 
$\vec{\mathcal{A}_0}$ are \citep{MeOg02},
\be \label{moduli_shear}
\hat{\mathcal{A}}_{0jilk} = 2 \widehat{W}_1 \delta_{i k} B_{j l}
 + 4 \widehat{W}_{11} B_{i j} B_{l k},
\en
where 
$\widehat{W}_{1} = \widehat{W}'(I_1)$, 
$\widehat{W}_{11} = \widehat{W}''(I_1)$.
With the help of \eqref{B_shear}$_2$, we find that 
in the ($\hat{x}_1, \hat{x}_2$) coordinate system,
the non-zero components $\hat{\mathcal{A}}_{0jilk}$ relevant 
to in-plane motion are
\begin{align} \label{A0_K}
& 
\hat{\mathcal{A}}_{01112} = \hat{\mathcal{A}}_{01211} = 
  4 \widehat{W}_{11} K(1+K^2), &&
\notag \\
& \hat{\mathcal{A}}_{01121} = \hat{\mathcal{A}}_{02111} = 
  2 \widehat{W}_{1} K + 4 \widehat{W}_{11} K(1+K^2), &&
\notag \\
& \hat{\mathcal{A}}_{01122} = \hat{\mathcal{A}}_{02211} = 
  4 \widehat{W}_{11} (1+K^2),
&&
\hat{\mathcal{A}}_{01221} = \hat{\mathcal{A}}_{02112} = 
  4 \widehat{W}_{11} K^2,
\notag \\
& \hat{\mathcal{A}}_{01212} = 
  2 \widehat{W}_{1} (1+K^2) + 4 \widehat{W}_{11} K^2,
&&
\hat{\mathcal{A}}_{02121} = 
  2 \widehat{W}_{1} + 4 \widehat{W}_{11} K^2,
\notag \\
& \hat{\mathcal{A}}_{01222} = \hat{\mathcal{A}}_{02212} = 
  2 \widehat{W}_{1} K + 4 \widehat{W}_{11} K,
&&
\hat{\mathcal{A}}_{02122} = \hat{\mathcal{A}}_{02221} = 
  4 \widehat{W}_{11} K,
 \notag \\
& 
\hat{\mathcal{A}}_{01111} = 
 2 \widehat{W}_{1}(1+K^2) + 4\widehat{W}_{11}(1  + K^2)^2,
&&
\hat{\mathcal{A}}_{02222} =  2 \widehat{W}_{1} + 4\widehat{W}_{11}.
\end{align}

Now the remaining incremental governing equations 
\eqref{governing}$_{1,2,4}$ reduce to
\begin{equation} \label{motion1}
- S_{11} + \ii S'_{21} = - \rho v^2 U_1, 
\quad 
- S_{12} + \ii S'_{22} = - \rho v^2 U_2, 
\quad 
\ii U_1 + U'_2 = 0,
\end{equation}
where by \eqref{s},
\begin{align} \label{S_shear}
& \ii S_{11} = 
  \ii (\hat{\mathcal{A}}_{01111} + \hat{p}) U_1  
    + \ii \hat{\mathcal{A}}_{01112} U_2
      + \hat{\mathcal{A}}_{01121}U_1' 
        +  \hat{\mathcal{A}}_{01122} U_2' - \ii \hat{P},
\nonumber \\
& \ii S_{12} = 
    \ii \hat{\mathcal{A}}_{01112}U_1  
      + \ii  \hat{\mathcal{A}}_{01212} U_2
         + (\hat{\mathcal{A}}_{01221} + \hat{p})U_1' 
           + \hat{\mathcal{A}}_{01222} U_2'.
\nonumber \\
& \ii S_{21} = 
   \ii \hat{\mathcal{A}}_{01121}U_1  
    + \ii (\hat{\mathcal{A}}_{01221} + \hat{p}) U_2
      + \hat{\mathcal{A}}_{02121}  U_1' 
        +  \hat{\mathcal{A}}_{02122} U_2',
\nonumber \\
& \ii S_{22} = 
   \ii \hat{\mathcal{A}}_{01122}U_1  
     + \ii \hat{\mathcal{A}}_{01222} U_2
       + \hat{\mathcal{A}}_{02122}U_1' 
          +  (\hat{\mathcal{A}}_{02222} + \hat{p}) U_2' - \ii \hat{P},
\end{align}
where $\hat{P}$ is the increment of $\hat{p}$.
The pre-stress necessary to maintain the solid in the static 
state of large simple shear is \eqref{prestress_plane}.
In particular, the $\hat{\sigma}_{22}$ component along the 
$\hat{x}_2$ axis is found using \eqref{B_shear} and \eqref{What}
as
\be
\hat{\sigma}_{22} = -\hat{p} +   2\widehat{W}_1,  
\en
leading to the connection
\be
\hat{\mathcal{A}}_{01221}
  +  \hat{p} = \hat{\mathcal{A}}_{02121} - \hat{\sigma}_{22}.
\en
Note that  \citet{CoOg95}  and \citet{Fu05a} keep $\sigma_2$ rather than
$\hat{\sigma}_{22}$ as a measure of the pre-stress. 
It is the component of the pre-stress along the principal 
axis $x_2$, whose orientation changes with the pre-strain,
in contrast to $\hat{\sigma}_{22}$, the component of the Cauchy pre-stress tensor 
along the \emph{unchanged} normal to the interface.
As pointed out by \citet{HuOg00}, it is $\hat{\sigma}_{22}$ which 
is continuous across the interface of two bonded sheared solids.

Following the same procedure as in Section \ref{section_3_1}, we find the Stroh 
formulation of the governing equations in the expected form 
\eqref{governing}, where $\vec{\xi} = [U_1, U_2, S_{21}, S_{22}]^t$ 
and $-\vec{N}_1$ and $\vec{N}_2$, are given by
\be \label{N1N2_shear}
 \begin{bmatrix}
   K & 1 - \dfrac{\hat{\sigma}_{22}}{2 \widehat{W}_{1} + 4 \widehat{W}_{11} K^2}
        \\
        1 & 0
       \end{bmatrix}, \quad
  \begin{bmatrix}
       \dfrac{1}{2 \widehat{W}_{1} + 4 \widehat{W}_{11} K^2} & 0  \\
                         0 & 0
     \end{bmatrix},
\en
respectively, and $-\vec{N}_3$ by
\be \label{N3_shear}
 \begin{bmatrix}
     2(4 \widehat{W}_{1} - \hat{\sigma}_{22})
         &  -(4 \widehat{W}_{1} - \hat{\sigma}_{22})K
 \\
   -(4 \widehat{W}_{1} - \hat{\sigma}_{22})K
     &   2 \widehat{W}_{1}K^2 + 2\hat{\sigma}_{22}
           - \dfrac{\hat{\sigma}_{22}^2}{2 \widehat{W}_{1} + 4 \widehat{W}_{11} K^2}
                                     \end{bmatrix}.
\en

    \CCLsubsection{Resolution}

The \emph{propagation condition} \eqref{propagation_condition}
reduces to a quartic in $q$,
\begin{multline} \label{quartic_shear}
2 (\widehat{W}_{1} + 2 \widehat{W}_{11} K^2) q^4 
 + 4 (\widehat{W}_{1} + 2 \widehat{W}_{11} K^2) K q^3 \\
  + [2 (\widehat{W}_{1}(2+K^2) - 4 \widehat{W}_{11} K^2(2-K^2) - \rho v^2] q^2 \\
  + 4 (\widehat{W}_{1} - 2 \widehat{W}_{11} K^2)K  q \\
    + 2 \widehat{W}_{1}(1+K^2) + 4 \widehat{W}_{11} K^2  - \rho v^2 =0.
\end{multline}
Notice how $\hat{\sigma}_{22}$, though present in $\vec{N}$, does not appear 
explicitly in this equation.

Quite surprisingly, there are two instances -- both pointed out 
by  \cite{CoOg95, CoOg96} -- 
where we can solve this quartic exactly and simply.
The first instance occurs for incremental \emph{deformations}, 
when $v = 0$;
it is applicable whatever the strain energy density $W$ might be. 
The reason for the simplicity of this resolution is made apparent by the 
change of unknown from $q$ to $\widetilde{q} = q - K/2$.
Then the quartic becomes
\begin{multline} \label{reduced_shear}
2 (\widehat{W}_{1} + 2 \widehat{W}_{11} K^2) \widetilde{q}^4  
  + [\widehat{W}_{1}(4-K^2) - 2 \widehat{W}_{11} K^2(4+K^2) - \rho v^2] \widetilde{q}^2 \\
  + (\rho v^2)K \widetilde{q}  
    + [ (\widehat{W}_{1} + 2 \widehat{W}_{11} K^2)(4+K^2)  - 2\rho v^2](4+K^2)/8 =0,
\end{multline} 
which is clearly a biquadratic at $v = 0$.
The consequence is that any incremental \emph{static} problem
can be solved in its entirety for sheared solids because the roots $q$ are accessible
explicitly.
This was to be expected though, because of an important theorem by \citet{FuMi02} which states that 
``the buckling condition for a pre-stressed elastic half-space is independent
of the orientation of the surface as long as the surface normal remains in
the ($x_1, x_2$) plane''; 
the \emph{buckling condition} is what corresponds to the marginally stable 
static solution obtained at $v = 0$; 
it is also called the \emph{wrinkling condition},
or the \emph{bifurcation criterion}, or any other denomination associated with the 
onset of instability in the linearised (incremental) theory.

The second instance where the quartic is easy to solve is when the 
solid is a Mooney-Rivlin material, see \eqref{mooney};
this case is treated in Section \ref{section_4_3_1}.

In general however, the quartic is difficult (but not impossible, 
see \citet{Fu05a, Fu05b}, \citet{DeFu06}, and the concluding Section)
to solve analytically and other methods, such as those relying 
on the fundamental equations \eqref{fundamental}, are required.
For the time being, we complete the picture with formal calculations.

Assuming the roots of the quartic have been computed, and calling 
them $q_1$, $q_2$, $\overline{q}_1$, $\overline{q}_2$,
where $q_1$ and $q_2$ both satisfy the decay condition \eqref{decay},
we find that the eigenvectors $\vec{\zeta}^1$ and  $\vec{\zeta}^2$ 
associated with $q_1$ and $q_2$, respectively, are
\be
\vec{\zeta}^1 = \begin{bmatrix}\vec{a}^1 \\ \vec{b}^1\end{bmatrix}, \quad
\vec{\zeta}^2 = \begin{bmatrix}\vec{a}^2 \\ \vec{b}^2\end{bmatrix}, 
\en
where
\begin{align} \label{a_b_shear}
& \vec{a}^j = [q_j^2, -q_j]^t, 
\notag \\
& \vec{b}^j = \left[ q_j 
  \left(\hat{\gamma}_{21} (q_j^2 + K q_j - 1) + \hat{\sigma}_{22} \right),
  -(\hat{\gamma}_{21} -\hat{\sigma}_{22})q_j^2 + \hat{\nu}_{12}q_j
                 + \hat{\gamma}_{12} - \rho v^2 \right]^t. 
\end{align}
Here,
\begin{align} \label{gamma_shear}
 & \hat{\gamma}_{21} = 2 \widehat{W}_{1} + 4 \widehat{W}_{11}K^2, 
 && 
  \hat{\gamma}_{12} = 2 \widehat{W}_{1}(1+K^2) + 4 \widehat{W}_{11}K^2, 
\notag \\
& \hat{\beta}_{12} = \widehat{W}_{1}(2+K^2) - 2 \widehat{W}_{11}K^2(2 - K^2),
&&
\hat{\nu}_{12} = 2 \widehat{W}_{1}K - 4 \widehat{W}_{11}K^3. 
\end{align}
Then we compute the $\vec{A} = [\vec{a}^1 | \vec{a}^2]$ and 
$\vec{B} = [\vec{b}^1| \vec{b}^2]$ matrices, and eventually 
the surface impedance 
matrix $\vec{M} = - \ii \vec{B A}^{-1}$, as
\be 
\vec{M} = - \ii  
 \begin{bmatrix}
  \hat{\gamma}_{21}(K+q_1 + q_2) & \hat{\gamma}_{21}(1 + q_1 q_2) + \hat{\sigma}_{22} 
   \vspace{6pt} \\
  \dfrac{\rho v^2 - \hat{\gamma}_{12}}{q_1 q_2} - \hat{\gamma}_{21} - \hat{\sigma}_{22} & 
     (\rho v^2 - \hat{\gamma}_{12})\dfrac{q_1 + q_2}{q_1 q_2} - \hat{\nu}_{12}
  \end{bmatrix}.
\en
Note that this matrix is indeed Hermitian;
this is easy to show by using the quartic \eqref{quartic_shear} and the definitions 
\eqref{gamma_shear} to uncover the identities: 
\begin{align}
& q_1 +  q_2 + \overline{q}_1 + \overline{q}_2 = -2 K, \notag \\
& q_1 q_2 \overline{q}_1 \overline{q}_2 = (\hat{\gamma}_{12} - \rho v^2)/\hat{\gamma}_{21},
\notag \\
& q_1 q_2 (\overline{q}_1 + \overline{q}_2)
  +  \overline{q}_1 \overline{q}_2 ( q_1 + q_2) 
   = -2 \hat{\nu}_{12}/\hat{\gamma}_{21}.
\end{align}
These identities allow us to rewrite the surface impedance matrix as
\be \label{M_shear}
\vec{M} = - \ii  
 \begin{bmatrix}
  \hat{\gamma}_{21}(q_1 + q_2 - \overline{q}_1 - \overline{q}_2)/2 & \hat{\gamma}_{21}(1 + q_1 q_2) - \hat{\sigma}_{22} 
   \vspace{6pt} \\
  -\hat{\gamma}_{21}(1 + \overline{q}_1 \overline{q}_2) + \hat{\sigma}_{22}& 
     - \hat{\gamma}_{21}[q_1 q_2 (\overline{q}_1 + \overline{q}_2) 
                                   -\overline{q}_1 \overline{q}_2 (q_1 + q_2)]/2
  \end{bmatrix}.
\en

\citet{Fu05a} derived the same form of the surface impedance matrix for the more general case of 
\emph{any plane strain} (any pre-strain where $\lambda_3 = 1$), 
which includes the present case of finite shear.

Notice also how the pre-stress $\hat{\sigma}_{22}$ is going to be explicitly present 
in the secular equation for Rayleigh surface waves \eqref{secularM},
but absent from the secular equation for Stoneley interface waves \eqref{stoneley}.
These secular equations remain implicit as long as the roots $q_1$ and $q_2$ are not known.
In the general case where the quartic \eqref{quartic_shear} is not solvable in 
a simple manner, we seek an explicit secular equation using the fundamental 
equations \eqref{fundamental}.

\CCLsubsubsection{Rayleigh surface waves.}
\label{section_4_2_1}

With the explicit expressions \eqref{N1N2_shear} and 
\eqref{N3_shear} for the blocks of the matrix $\vec{N}$,
we can compute $\vec{N}^{-1}$ and $\vec{N}^2$. 
The lower left corner of these gives in turn
$\vec{K}^{(1)} = \vec{N}_3 + \rho v^2 \vec{I}$
and $\vec{K}^{(-1)}$, $\vec{K}^{(2)}$, respectively. 
Then the equations \eqref{fundamental_rayleigh} written at
$n=1,-1,2$, yield the linear homogeneous system
\begin{equation}\label{Kmat}
  \begin{bmatrix}
   K^{(1)}_{11}  & K^{(1)}_{12}  & K^{(1)}_{22}  \\
   K^{(-1)}_{11} & K^{(-1)}_{12} & K^{(-1)}_{22} \\
   K^{(2)}_{11}  & K^{(2)}_{12}  & K^{(2)}_{22}
 \end{bmatrix}
  \begin{bmatrix}
    U_1(0) \overline{U}_1(0) \\
    U_1(0) \overline{U}_2(0) +  \overline{U}_1(0) U_2(0) \\
    U_2(0) \overline{U}_2(0)
  \end{bmatrix}
= \begin{bmatrix} 0 \\ 0 \\ 0 \end{bmatrix}.
\end{equation}
The vanishing of the determinant of the $3 \times 3$ matrix on the 
left hand side is the explicit polynomial secular equation for surface waves
in a sheared incompressible semi-infinite solid. 
It is a polynomial of degree 4 in $\rho v^2$. 
It is too long to reproduce in general but easy to obtain (and 
solve numerically) with a computer algebra system. 
Here we present its expression in the case where $\hat{\sigma}_{22} = 0$.
Then,  $\vec{K}^{(1)}$ and $ \vec{K}^{(2)}$ are given by
\be \label{K_1_2_shear}
   \begin{bmatrix}
     \rho v^2 - 8 \widehat{W}_{1} &  4\widehat{W}_{1} \\
     4 \widehat{W}_{1}            &  \rho v^2 + 2 \widehat{W}_{1}
                  \end{bmatrix},
\quad
  \begin{bmatrix}
     2(4 \widehat{W}_{1} - \rho v^2)K &  2\widehat{W}_{1}(4 - K^2) - 2 \rho v^2 \\
     2\widehat{W}_{1}(4 - K^2) - 2 \rho v^2 & - 8\widehat{W}_{1}K
                  \end{bmatrix},
 \en
respectively, the components of $\vec{K}^{(-1)}$ are
\begin{align} \label{K_minus1_shear}
&  K^{(-1)}_{11} = 2(2 \widehat{W}_{1} - \rho v^2), 
\notag \\
&  K^{(-1)}_{12} = -2[2 \widehat{W}_{1}(1+K^2) - \rho v^2]K, 
\notag \\
&  K^{(-1)}_{22} = 4 \widehat{W}_{1}(2+K^2)^2 
  + \rho v^2\dfrac{\rho v^2 - 2\widehat{W}_{1}(5+K^2) - 4 \widehat{W}_{11}K^2(1+K^2)}
        {\widehat{W}_{1} + 2 \widehat{W}_{11} K^2}, 
\end{align}
(up to an inessential common factor), and the secular equation is the quartic
\begin{multline} \label{quartic_secular}
x^4 - 5x^3 + 
 \left( 8 \dfrac{5+K^2}{4+K^2} 
        + K^2\dfrac{\widehat{W}_{11}}{\widehat{W}_1}\right) x^2
 \\
-8  \left( 4 \dfrac{1+K^2}{4+K^2} 
        + K^2\dfrac{\widehat{W}_{11}}{\widehat{W}_1}\right) x
 + 8 \dfrac{\widehat{W}_{1} + 2 \widehat{W}_{11} K^2} 
            {\widehat{W}_{1}(4+K^2)} = 0,
\end{multline}
where $x$ is the following non-dimensional measure of the squared wave speed,
$x = \rho v^2/[\widehat{W}_{1}(4 + K^2)]$.

As stated above, the secular equation is also a quartic in the squared wave 
speed when $\hat{\sigma}_{22} \ne 0$. 
For a given material, a given pre-stress, and a given shear, its numerical resolution may 
yield more than one positive real root. 
If such is the case, then for each corresponding speed, 
compute the roots to the quartic \eqref{quartic_shear} and discard 
those (supersonic) speeds which do not give two complex conjugate pairs of roots. 
Finally, find which of the remaining speeds (if there is more than one) 
satisfies the secular equation written in optimal form \eqref{secularM}.

\CCLsubsubsection{Stoneley shear-twin interface waves.}

For Stoneley interface waves, the fundamental 
equations \eqref{fundamental} are not practical to derive secular 
equations in general, and we must resort to other methods, such as 
those developed by \cite{DeFu06}. 
The exception is the special case when each half-space is filled with Mooney-Rivlin 
materials, because then the roots to the quartic \eqref{quartic_shear} can be found easily,
see Section \ref{section_4_3_1}.
 
\begin{figure}  [htbp]
\centering 
 \epsfig{figure=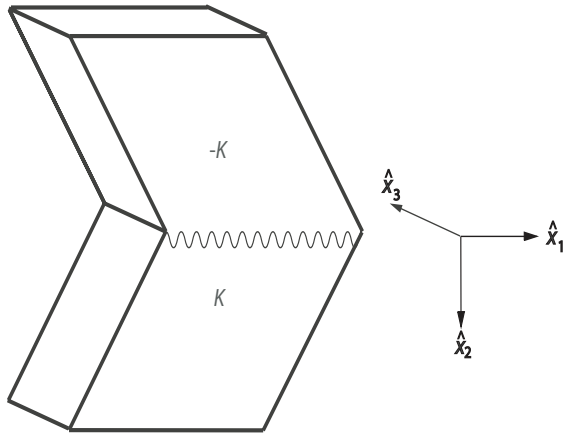, width=.55\textwidth, height=0.3\textheight}
 \caption{
 Stoneley wave propagation at a shear-twin interface. 
 Both half-spaces are occupied with the same incompressible 
 solid, one subject to a simple shear of amount $K$, the 
 other subject to a simple shear of amount $-K$.
 The analysis shows that Stoneley waves cannot actually travel in the direction 
 of shear. \label{fig_5}}
\end{figure}

We now focus on the possibility of propagating incremental 
waves at a \emph{shear-twin interface}. 
In this configuration, two solids, made of the same incompressible
material, are subject to equal and opposite shears, see Figure \ref{fig_5}. 
The study of wave propagation at this type of interface can have important repercussions 
in the non-destructive evaluation of a twinned interface because this bimaterial can 
``simulate the finite (plastic) deformation associated with a crystal twin''
\citep{HuOg00}.

For one half-space, the propagation condition is the quartic \eqref{quartic_shear}.
For the other half-space, the amount of shear is changed to its opposite,
but $\widehat{W}_{1}$ and  $\widehat{W}_{11}$ do not change 
signs because they are functions of $K^2$.
It follows that if $q_1$ and $q_2$ are qualifying roots of the 
propagation condition in one half-space, then $-q_1$ and $-q_2$ are 
qualifying roots in the other half-space. 
Also, we see from  \eqref{gamma_shear} that 
$\hat{\gamma}_{21}$, $\hat{\gamma}_{12}$, and $\hat{\beta}_{21}$ in one half-space 
are equal to their counterparts in the other half-space, but that the counterpart to 
$\hat{\nu}_{12}$ is $-\hat{\nu}_{12}$.
Finally, $\hat{\sigma}_{22}$ is continuous across the interface.
Then we conclude from \eqref{a_b_shear} that the counterparts to the matrices 
\be 
\vec{A} = \begin{bmatrix} A_{11} & A_{12} \\ A_{21} & A_{22} \end{bmatrix}, \qquad
\vec{B} = \begin{bmatrix} B_{11} & B_{12} \\ B_{21} & B_{22} \end{bmatrix}, 
\en
in one half-space are the matrices \citep{Dest03, Ting05}
\be 
\vec{A}^* = \begin{bmatrix} A_{11} & A_{12} \\ -A_{21} & -A_{22} \end{bmatrix}, \qquad
\vec{B}^* = \begin{bmatrix} -B_{11} & -B_{12} \\ B_{21} & B_{22} \end{bmatrix}, 
\en
in the other half-space.
In turn, this conclusion leads to the following sum of surface impedance matrices 
\be \label{M_shear_twin}
\vec{M} +  \vec{M}^* =   
 \begin{bmatrix}
  2 M_{11} & 0 
   \vspace{6pt} \\
  0 &  2M_{22}
  \end{bmatrix},
\en
where $\vec{M}^* = \ii \vec{B}^* (\vec{A}^*)^{-1}$. 
Consequently, the exact secular equation for Stoneley shear-twin interface waves is,
according to \eqref{stoneley}, \eqref{M_shear}, and \eqref{M_shear_twin} that
either 
\be
 \Im (q_1 + q_2) = 0, \quad \text{or} \quad 
 \Im [q_1 q_2(\overline{q}_1 + \overline{q}_2)]=0.
\en
However, as is easily proved, neither of these quantities can be zero when 
both $q_1$ and $q_2$ have positive imaginary parts. 
It follows that \emph{Stoneley waves cannot propagate in the direction of sheared
at a shear-twin interface},
whatever the strain energy function is, and whatever what the shear is. 
Note however that \citet{HuOg00} show, in their study of reflection and transmission of plane
waves at shear-twin interface, that an incident harmonic plane wave can give rise to 
an interfacial wave (but of a different type than the Stoneley type).

    \CCLsubsection{Examples}

\CCLsubsubsection{Sheared Mooney-Rivlin solids.}
\label{section_4_3_1}

For the Mooney-Rivlin strain energy density \eqref{mooney} we have
\be
\widehat{W}_{1} = \mathcal{D}/2, \qquad 
\widehat{W}_{11} = 0,
\en
where $\mathcal{D} = \mathcal{D}_1 + \mathcal{D}_2$ is the shear modulus.
The quartic propagation condition \eqref{quartic_shear} then factorizes to
\begin{equation} \label{factorize_shear}
(q^2+1) (q^2 + 2 K q + K^2 + \eta^2) = 0,
\quad \text{where} \quad 
  \eta=  \sqrt{1 - \rho v^2/ \mathcal{D}},
\end{equation}
and its roots with positive imaginary parts are
\begin{equation} \label{q1q2_shear_mr}
  q_1= \ii, \qquad q_2= -K + \ii \eta.
\end{equation}
We also have 
\be
\hat{\gamma}_{21} = \mathcal{D}, 
 \quad 
  \hat{\gamma}_{12} = \mathcal{D}(1+K^2), 
\quad
 \hat{\beta}_{12} = \mathcal{D}(2+K^2),
\quad
\hat{\nu}_{12} = \mathcal{D}K, 
\en
and the surface impedance matrix of \eqref{M_shear} reduces to 
\be \label{M_mr_shear}
\vec{M} = \mathcal{D} 
 \begin{bmatrix}
  \eta+1 & -K  + \ii(\eta-1) + \ii \hat{\sigma}_{22}/\mathcal{D}  
   \vspace{6pt} \\
   -K - \ii(\eta-1) - \ii \hat{\sigma}_{22}/\mathcal{D}  & 
    \eta^2 + \eta + K^2
  \end{bmatrix}.
\en

\begin{figure}  [htbp]
\centering 
 \epsfig{figure=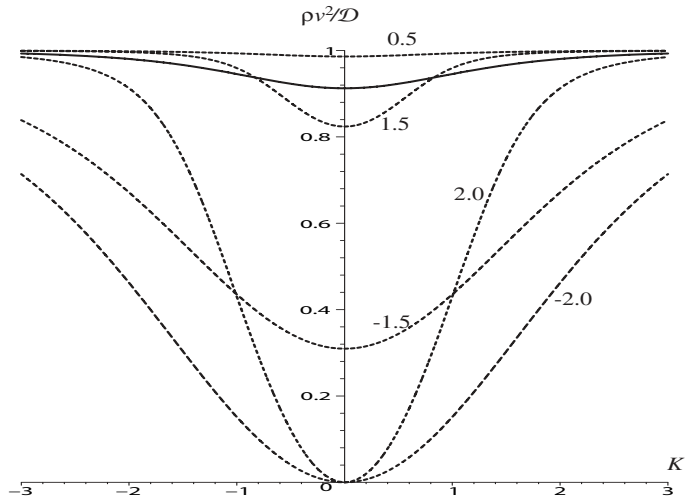, width=.75\textwidth, height=0.35\textheight}
 \caption{
 Squared scaled surface wave speed $\rho v^2 / \mathcal{D}$
 as a function of the amount of shear $K$ for a Mooney-Rivlin material.
 For the solid plot, there is no pre-stress normal to the boundary 
 ($\hat{\sigma}_{22}=0$);
 for the dashed plots, the displayed number indicates the value of the non-dimensional 
 pre-stress $\hat{\sigma}_{22} /\mathcal{D}$.
 More plots are found in the article by  \cite{CoOg95}.
 \label{fig_mr_shear}}
\end{figure}

For Rayleigh surface waves in a sheared Mooney-Rivlin material, 
the secular equation \eqref{secularM} is a cubic in $\eta$, 
\be \label{secular_mr_shear}
\eta^3 + \eta^2 + (3+ K^2 - 2 \hat{\sigma}_{22}/\mathcal{D})\eta
 - (1 - \hat{\sigma}_{22}/\mathcal{D})^2 = 0.
\en
See  \cite{CoOg95} for this equation with $\sigma_2$ instead 
of $\hat{\sigma}_{22}$, and \cite{DeOg05} for a generalization 
of this equation to a triaxial stretch followed by a shear.
See also Figure \ref{fig_mr_shear} for the variations of the squared scaled wave speed 
$\rho v^2 / \mathcal{D} = 1- \eta^2$ 
with the amount of shear $K$, for several values of the pre-stress $\hat{\sigma}_{22}$.
In particular, the plots at $\hat{\sigma}_{22} = \pm 2\mathcal{D}$ show 
that the Mooney-Rivlin material is unstable when subject only to a 
hydrostatic pressure of that amount (that is, $\rho v^2 = 0$ at $K = 0$) 
but regains stability as soon as it is sheared ($\rho v^2 > 0$ at $K \ne 0$) . 

As is clear from the roots \eqref{q1q2_shear_mr}, from 
the secular equation \eqref{secular_mr_shear}, and from Figure \ref{fig_mr_shear},
both the penetration depth and the speed are even functions of $K$.
Also, once $\eta$ is found by solving the secular equation, we 
have $\partial \eta / \partial (K^2) =
 -\eta^2 / [2\eta^3 + \eta^2 + (1- \hat{\sigma}_{22}/\mathcal{D})^2]<0$
and 
$\partial \Im (q_2) / \partial (K^2) =
 \partial \eta / \partial (K^2)<0$,
indicating that both the surface wave speed and the penetration depth 
increase with the magnitude of the shear.

We note that having found an explicit expression \eqref{M_mr_shear}
for the surface impedance matrix allows us to solve 
the problem of Stoneley interface waves in its 
entirety, for half-spaces made of different Mooney-Rivlin solids, subject to 
different amounts of shear;
see \cite{ChJa79b} for more discussion on this point.

\CCLsubsubsection{Sheared Gent solids.}

As an example of a strain energy density for which the quartic \eqref{quartic_shear} is 
not easily solved, we now work with the following function,
\begin{equation} \label{Gent}
W = -\textstyle{\frac{1}{2}}
  \mathcal{C} J_m
   \ln \left(1 -
    \dfrac{I_1 - 3}{J_m} \right),
\qquad I_1 < 3 + J_m,
\end{equation}
proposed initially by \citet{Gent96} to describe strain-stiffening elastomers
and since then extensively used by \citet[and references therein]{HoSa03}
to model strain-stiffening soft biological tissues such as arteries. 
Here $\mathcal{C}>0$ is the infinitesimal shear modulus and 
$J_m >0$ is a constant, accounting for the limiting chain extensibility of the solid.
Hence the condition \eqref{Gent}$_2$ limits the amount of shear by imposing 
\begin{equation} \label{gent_max}
-\sqrt{J_m} < K < \sqrt{J_m}.
\end{equation}

Turning our attention to the propagation of surface waves, we compute the following 
quantities,
\begin{equation}
\widehat{W}_{1} = \dfrac{\mathcal{C} J_m}{2(J_m - K^2)},
\qquad 
\widehat{W}_{11} = \dfrac{\mathcal{C} J_m}{2(J_m - K^2)^2}.
\end{equation}

\begin{figure}  [htbp]
\centering 
 \epsfig{figure=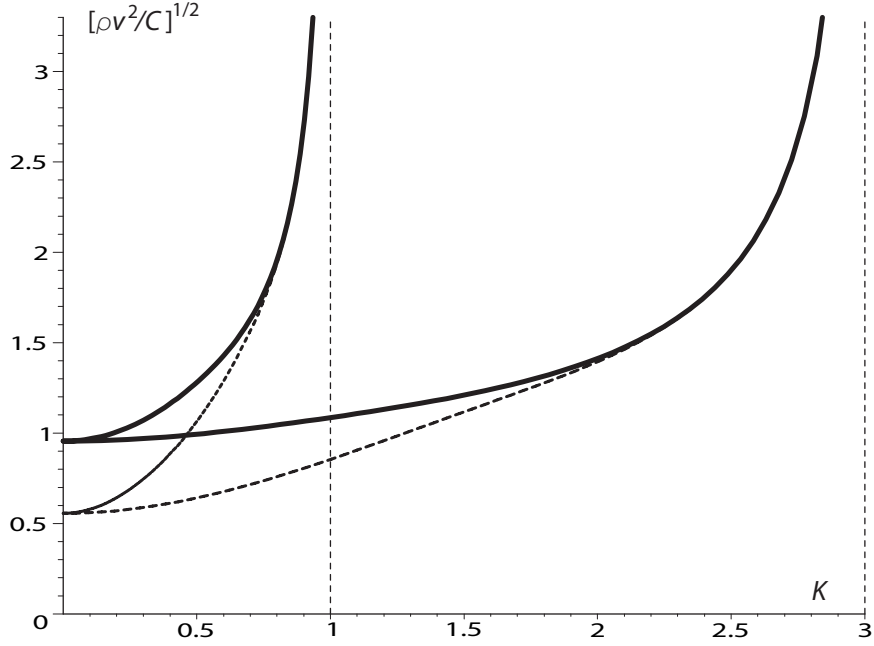, width=.85\textwidth, height=0.45\textheight}
 \caption{
 Scaled surface wave speed $\sqrt{\rho v^2 / \mathcal{C}}$
 as a function of the amount of shear $K$ for two Gent materials.
 For the solid plots, there is no pre-stress normal to the boundary 
 ($\hat{\sigma}_{22}=0$);
 for the dashed plots, the pre-stress is $\hat{\sigma}_{22} = -1.5 \mathcal{C}$.
 The vertical asymptotes correspond to the maximal amount of shear, that is
 $K_{\text{max}} = \pm 3.0$ for 
 the Gent solid with $J_m = 9.0$, and $K_{\text{max}} = \pm 1.0$ 
  for the other, stiffer, Gent solid with $J_m = 1.0$.
  \label{fig_gent_shear}}
\end{figure}

To fix the ideas, we take two Gent materials, one with 
$J_m = 9.0$, the other (stiffer) with $J_m = 1.0$.
Figure \ref{fig_gent_shear} shows the variations of the non-dimensional measure of the 
surface wave speed $\sqrt{\rho v^2 / \mathcal{C}}$ with the amount of shear $K$.
The bounds due to the inequalities \eqref{gent_max} are clearly visible, indicating 
that the solids become more and more rigid as their limit of chain extensibility is approached.
The speed is an even function of $K$ and only the $K \ge 0$ needs be displayed. 
For a boundary free of pre-stress ($\hat{\sigma}_{22}=0$), 
we solve numerically the quartic secular equation \eqref{quartic_secular} and find 
in general that there is two real positive roots for $\rho v^2$; 
one gives a supersonic speed and is discarded; 
the other gives a speed for which the exact secular equation \eqref{secularM} is satisfied and 
it is kept.
The Figure also shows (dotted plots) 
the effect of the compressive pre-stress $\hat{\sigma}_{22} = -1.5 \mathcal{C}$,
which is to slow the wave down in the small-to-moderate shear region;
here again a quartic secular equation is solved numerically and only one speed is kept. 

    \CCLsection{Propagation in a principal plane}
    \label{section_5}
   
\begin{figure}[htbp]
\centering 
 \epsfig{figure=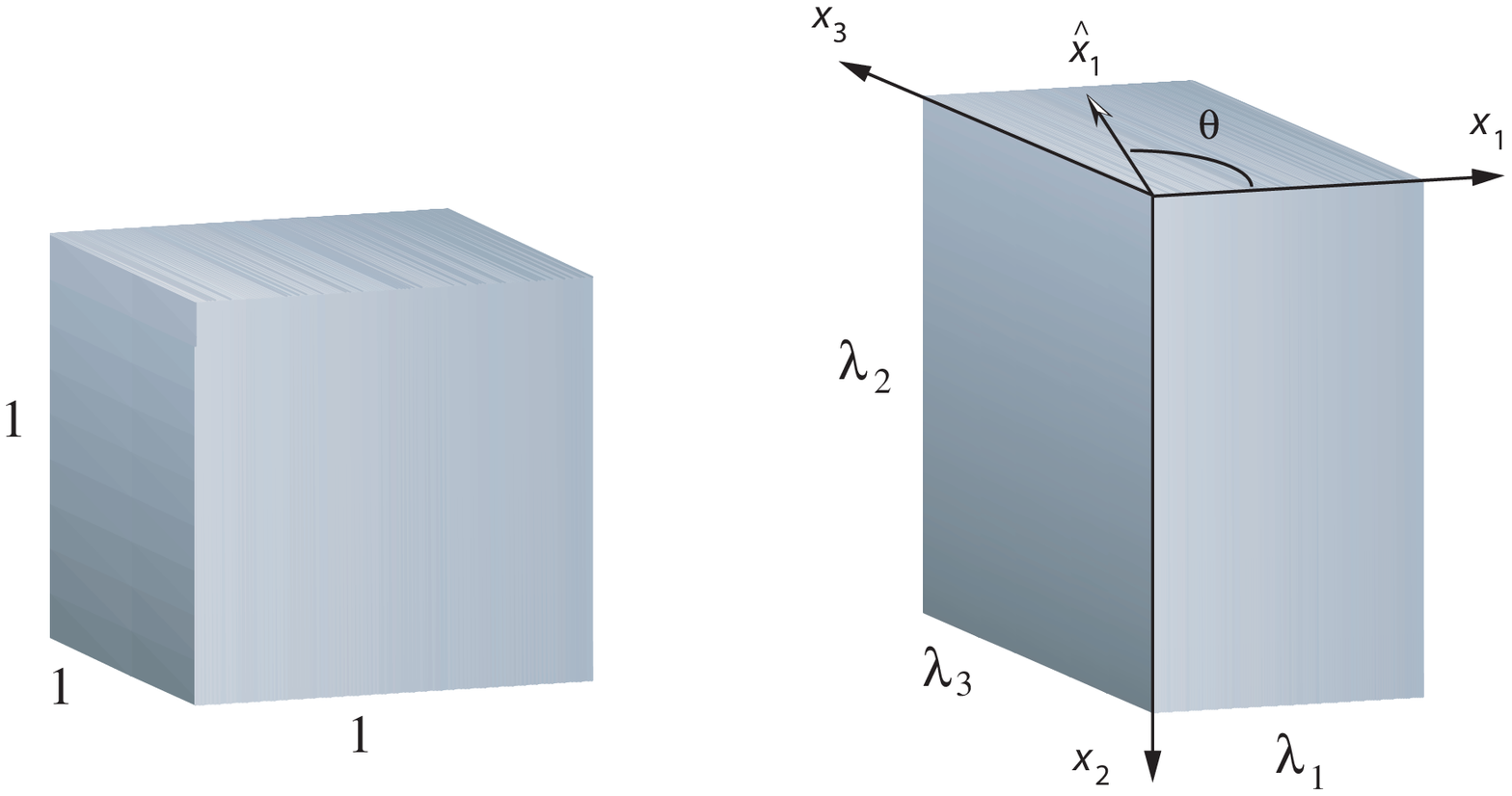, width=.9\textwidth}
 \caption{
 Interface wave propagation when the boundary $x_2 = 0$
 is a principal plane of pre-strain. 
 The $\hat{x}_1$ direction is the direction of propagation, making an angle $\theta$
 with the principal direction of pre-strain $x_1$.
  \label{fig_non_principal}}
\end{figure}

In this Section we consider an interface wave propagating in a principal plane, 
$x_2 = 0$ say, but not in a principal direction. 
We call $\theta$ the angle between the propagation direction $\hat{x}_1$ and the principal 
axis $x_1$, see Figure \ref{fig_non_principal}.
Although some results exist for non-principal Stoneley waves \citep{Dest05}, 
the focus of this Section is on Rayleigh surface waves.
 
    \CCLsubsection{Governing equations}
   
Hence we model this motion as
\begin{equation} \label{wave}
\{ \vec{u}, \dot{p}, \vec{s} \}
  = \{ \vec{U}(kx_2), \ii kP(k x_2),
          \ii k\vec{S}(k x_2)\}
                \ee^{\ii k(c_\theta x_1 + s_\theta x_3 - v t)},
\end{equation}
where $c_\theta = \cos\theta$, $s_\theta = \sin\theta$.
By projecting the governing equations in the coordinate axis of the principal 
axes (where $\vec{\mathcal{A}_0}$ has 15 independent non-zero components, 
see Section \ref{2_2}), we can write them in the Stroh form  \eqref{motion} as a
homogeneous linear system of six first-order differential equations,
where $\vec{\xi} =
    [U_1, U_2, U_3, S_{21}, S_{22}, S_{23}]^t$,
see \cite{DOPR05} for details.
Here the matrices $-\vec{N}_1$, $\vec{N}_2$, $-\vec{N}_3$ are given by
\begin{equation}  \label{N_non_principal}
 \begin{bmatrix}
       0 & c_\theta (\gamma_{21}-\sigma_2)/\gamma_{21} & 0 \\
       c_\theta & 0 & s_\theta \\
       0 & s_\theta (\gamma_{23}-\sigma_2)/\gamma_{23} & 0
       \end{bmatrix},
\quad
 \begin{bmatrix}
       1/\gamma_{21} & 0 & 0 \\
           0 & 0 & 0 \\
        0 & 0 &  1/\gamma_{23}
      \end{bmatrix},
\quad
  \begin{bmatrix}
                       \chi   &     0     & -\kappa \\
                          0   &     \nu   &      0     \\
                    - \kappa  &     0     &  \mu
                   \end{bmatrix},
\end{equation}
respectively, where
\begin{align}
& \chi =
    2 c_\theta^2(\beta_{12} + \gamma_{21} - \sigma_2)
          + s_\theta^2 \gamma_{31},
\nonumber \\
& \nu =
  c_\theta^2[\gamma_{12}
      - (\gamma_{21}-\sigma_2)^2 / \gamma_{21}]
   + s_\theta^2[\gamma_{32}
            - (\gamma_{23}-\sigma_2)^2/\gamma_{23}],
\nonumber \\
& \mu =
    c_\theta^2 \gamma_{13}
       +  2 s_\theta^2 (\beta_{23} + \gamma_{23} - \sigma_2),
\nonumber \\
& \kappa
    =   c_\theta s_\theta (\beta_{13} - \beta_{12} - \beta _{23}
                             - \gamma_{21} - \gamma_{23} + 2\sigma_2),
\end{align}
and the $\gamma_{i j}$, $\beta_{i j}$ are defined in  \eqref{gammaBeta}.

\citet{RoSa99} show that the \emph{propagation condition} 
\eqref{propagation_condition} is a cubic in $q^2$, 
\begin{equation} \label{bicubic}
\gamma_{21} \gamma_{23} q^6
 - [(\gamma_{21} + \gamma_{23})X - c_1]\,q^4
  + (X^2 - c_2X + c_3)\,q^2
    + (X-c_4)(X-c_5) = 0,
\end{equation}
with $X = \rho v^2$ and
\begin{align} \label{c_4}
& c_1 =  (\gamma_{21}\gamma_{13} + 2\beta_{12}\gamma_{23})c_\theta^2
        + (\gamma_{23}\gamma_{31} + 2\beta_{23}\gamma_{21})s_\theta^2,
\nonumber \\
& c_2 =  (\gamma_{23} + \gamma_{13} + 2\beta_{12})c_\theta^2
              + (\gamma_{21} + \gamma_{31} + 2\beta_{23})s_\theta^2,
\nonumber \\
& c_3 =  (\gamma_{12}\gamma_{23} + 2\beta_{12}\gamma_{13})c_\theta^4
        + (\gamma_{21}\gamma_{32} + 2\beta_{23}\gamma_{31})s_\theta^4
\nonumber \\
& \phantom{123456}  + [\gamma_{12}\gamma_{21} + \gamma_{13}\gamma_{31}
                 +\gamma_{23}\gamma_{32}
                   - (\beta_{13} - \beta_{12} - \beta_{23})^2
                    + 4 \beta_{12}\beta_{23}]c_\theta^2 s_\theta^2,
\nonumber \\
& c_4 = \gamma_{12}c_\theta^2 + \gamma_{32}s_\theta^2,
\nonumber \\
& c_5 =  \gamma_{13}c_\theta^4 + 2 \beta_{13}c_\theta^2s_\theta^2
                   + \gamma_{31}s_\theta^4.
\end{align}

    \CCLsubsection{Resolution for Rayleigh surface waves}

Finding the eigenvectors $\vec{\zeta}^1$, $\vec{\zeta}^2$, $\vec{\zeta}^3$ of $\vec{N}$ 
corresponding to the qualifying roots $q_1$, $q_2$, $q_3$ (say)
is a lengthy task by hand, best left to a computer
algebra system. 
In the end, we find that the $\vec{\zeta}^i$ are of the forms:
\be
\vec{\zeta}^1 = \begin{bmatrix} \vec{a}^1 \\
                  \vec{b}^1\end{bmatrix}, \qquad
\vec{\zeta}^2 = \begin{bmatrix} \vec{a}^2 \\ \vec{b}^2\end{bmatrix}, \qquad
\vec{\zeta}^3 = \begin{bmatrix} \vec{a}^3 \\ \vec{b}^3\end{bmatrix}, 
\en
where
\be
\vec{a}^j =\begin{bmatrix}
   a_4 q_i^4 + a_2 q_i^2 + a_0 \\
   -q_j^5 + b_3 q_j^3 + b_1 q_j \\
  d_4 q_j^4 + d_2 q_j^2 + d_0 
 \end{bmatrix}, 
\qquad
\vec{b}^j = \begin{bmatrix}
  h_3 q_j^3 + h_1 q_j \\
  (\nu - X)(q_j^4 + mq_j^2 + n) \\
  g_3 q_j^3 + g_1 q_j
 \end{bmatrix}.
\en
Here the quantities $m$ and $n$ are given by 
\begin{multline}
 m = \left(\frac{1}{\gamma_{21}}  +
            \frac{(\gamma_{21} - \sigma_2)^2}
             {\gamma_{21}^2(\nu -X)}c_\theta^2\right) [\eta-X]
 + \left(\frac{1}{\gamma_{23}}
 +\frac{(\gamma_{23}-\sigma_2)^2}{\gamma_{23}^2(\nu -X)}
                       s_\theta^2\right)[\mu-X]
\\ - 2\kappa\frac{(\gamma_{21}-\sigma_2)(\gamma_{23}-\sigma_2)}
                                    {\gamma_{21}\gamma_{23}(\nu -X)}
                                             c_\theta s_\theta,
\end{multline}
\begin{multline}
  n = \left\{ 1+
  \left[\frac{(\gamma_{21}-\sigma_2)^2}{\gamma_{21}}c_\theta^2
       +\frac{(\gamma_{23}-\sigma_2)^2}{\gamma_{23}}s_\theta^2\right]
                                                    (\nu -X)^{-1}
  \right\}
 \\
    \times [(\mu -X)(\eta-X)-\kappa^2]/(\gamma_{21}\gamma_{23}).
\end{multline}
The expressions for the quantities $a_i$, $b_i$, $d_i$, $h_i$,
$g_i$ are too lengthy to reproduce but they are easily obtained in a formal 
manner using a computer algebra system;
as it turns out, these constants are not needed in the secular equation
for Rayleigh surface waves.
Indeed, the expression for the surface impedance matrix $\vec{M}$ is 
very lengthy, but its determinant factorizes greatly. 
Hence we find that the exact secular equation 
for Rayleigh surface waves \eqref{secularM} 
reduces to \citep{Tazi87, DOPR05}
\begin{equation} \label{secularImplicit}
n \omega_\text{I} - \omega_\text{III}(m-\omega_\text{II}) = 0,
\end{equation}
where
\begin{equation} \label{omegas}
\omega_\text{I} = -(q_1 + q_2 + q_3),
\quad
\omega_\text{II} = q_1 q_2 + q_2 q_3 + q_3 q_1,
\quad
\omega_\text{III} = - q_1 q_2 q_3.
\end{equation}
This secular equation remains implicit as long as the roots $q_1$, 
$q_2$, $q_3$ satisfying the decay condition are not known.
To compute them, we must find the wave speed, by using the fundamental equations 
\eqref{fundamental_rayleigh}.

First, using the explicit expression \eqref{N_non_principal} of the Stroh 
matrix $\vec{N}$, we compute $\vec{N}^{-1}$ and $\vec{N}^3$, and in particular
we find explicit expressions for the lower left blocks
$\vec{K}^{(1)} = \vec{N}_3 + \rho v^2 \vec{I}$, $\vec{K}^{(-1)}$,
$\vec{K}^{(3)}$.
Next, we use the result \citep{Dest05} that $\vec{U}(0)$ is in the form
\begin{equation}
  \vec{U}(0)=U_{1}(0)[1,\ii \alpha, \beta]^\text{T},
\end{equation}
where $\alpha$, $\beta$ are \emph{real} numbers, 
to write the fundamental equations \eqref{fundamental_rayleigh}
at $n = -1, 1, 3$ as the following system of three equations,
\begin{equation} \label{systemK}
\begin{bmatrix}
K_{13}^{(-1)} & K_{33}^{(-1)} & K_{22}^{(-1)} \\
K_{13}^{(1)} & K_{33}^{(1)} & K_{22}^{(1)} \\
K_{13}^{(3)} & K_{33}^{(3)} & K_{22}^{(3)} \end{bmatrix}
\begin{bmatrix} 2 \beta\\
\beta^{2}\\
\alpha^{2}\end{bmatrix}
=
\begin{bmatrix}-K_{11}^{(-1)}\\
 -K_{11}^{(1)}\\
-K_{11}^{(3)}\end{bmatrix}.
\end{equation}
Then, we solve this non-homogeneous system by Cramer's rule to 
find 
\be \label{cramer}
2 \beta=\Delta_{1}/\Delta,
\qquad
\beta^{2}=\Delta_{2}/\Delta,
\en
where  $\Delta$, $\Delta_{1}$, and $\Delta_{2}$ are the following determinants, 
\begin{multline} \label{def_Delta}
\Delta=
\begin{vmatrix}
K_{13}^{(-1)} & K_{33}^{(-1)} & K_{22}^{(-1)} \\
K_{13}^{(1)} & K_{33}^{(1)} & K_{22}^{(1)} \\
K_{13}^{(3)} & K_{33}^{(3)} & K_{22}^{(3)}
\end{vmatrix},
\\
\Delta_{1}=
\begin{vmatrix}
-K_{11}^{(-1)} & K_{33}^{(-1)} & K_{22}^{(-1)} \\
-K_{11}^{(1)}  & K_{33}^{(1)}  & K_{22}^{(1)}  \\
-K_{11}^{(3)}  & K_{33}^{(3)}  & K_{22}^{(3)}
\end{vmatrix},\qquad\quad\;
\\
\Delta_{2}=
\begin{vmatrix}
K_{13}^{(-1)} & -K_{11}^{(-1)} & K_{22}^{(-1)} \\
K_{13}^{(1)}  & -K_{11}^{(1)}  & K_{22}^{(1)} \\
K_{13}^{(3)}  & -K_{11}^{(3)}  & K_{22}^{(3)}
\end{vmatrix}.
\end{multline}
Finally we write down the compatibility 
of equations \eqref{cramer} as
\begin{equation} \label{expl_secular}
\Delta_{1}^{2}-4\Delta\Delta_{2}=0,
\end{equation}
which is the \textit{explicit secular equation for non-principal
surface waves in deformed incompressible materials}.

In general this equation is a polynomial of degree 12 in $\rho v^2$ \citep{Tazi89, DOPR05}, 
easy to solve numerically. 
Of the 12 possible roots, we keep those which are real, positive, and 
give a subsonic speed (that is a speed for which the bicubic 
\eqref{bicubic} has three 
pairs of complex conjugate roots). 
We then test the remaining speeds against the exact secular equation \eqref{secularImplicit}.

    \CCLsubsection{Examples}

\CCLsubsubsection{neo-Hookean solid.}
 \label{section_5_3_1}

The neo-Hookean form of the strain energy density for an incompressible isotropic solid
is a sub-case of the Mooney-Rivlin form, namely $ \mathcal{D}_2 = 0$ in \eqref{mooney}
so that 
\be \label{neo}
W = \mathcal{D}_1 (\lambda_1^2 + \lambda_2^2 + \lambda_3^2)/2.
\en
It leads to a stress-strain relationship \eqref{prestress_I} which 
is ``linear'' with respect to the left Cauchy-green strain tensor,
\be 
\vec{\sigma} = -p \vec{I} + \mathcal{D}_1  \vec{B}.
\en
The neo-Hookean model is unable to capture neither qualitatively nor
quantitatively experimental data, over any range of deformations  \citep{Sacco04}. 
It is nonetheless a highly popular model in the literature because
it forms the basis of a statistical treatment for the molecular description of 
rubber elasticity \citep{Trel49}. 

With respect to wave propagation, it has more peculiar properties than 
the Mooney-Rivlin material, because here, 
\begin{equation}
\gamma_{ij}
  = \mathcal{D}_1 \lambda_i^2,
\qquad
      2\beta_{ij} =
 \mathcal{D}_1 (\lambda_i^2+\lambda_j^2),
\end{equation}
which leads to great simplifications in the matrices 
$\vec{N}_1$, $\vec{N}_2$, $\vec{N}_3$ of \eqref{N_non_principal}.

Now placing ourselves in the coordinates system ($\hat{x}_1, x_2, \hat{x}_3$)
attached to the wave propagation, see Figure \ref{fig_non_principal}, we introduce the functions
$\hat{U}_i$, $\hat{S}_{2i}$ ($i=1, 2, 3$), defined by
\begin{equation}
 \hat{U}_i = \Omega_{i j}U_j, \quad
 \hat{S}_{2i} = \Omega_{ij} S_{2j}, \quad
 \text{where} \quad
 \Omega_{ij} = \begin{bmatrix}
                    c_\theta & 0 & -s_\theta \\
                    0 & 1 & 0 \\
                   s_\theta & 0 & c_\theta
                \end{bmatrix}.
\end{equation}
With these functions, the governing equations \eqref{motion} decouple the
anti-saggital motion $[\hat{U}_3, \hat{S}_{23}]$ from its saggital
counterpart (recall that the direction of propagation and the normal to the interface
define what is called the \emph{saggital plane}.) 
For the latter motion we find
\begin{equation}
[\hat{U}_1', \hat{U}_2', \hat{S}_{21}', \hat{S}_{22}']^t
 =  \ii \vec{N} [\hat{U}_1, \hat{U}_2, \hat{S}_{21}, \hat{S}_{22}]^t,
\end{equation}
with 
\begin{equation} \label{Nprincipal}
\vec{N} = \begin{bmatrix}
  0\ &\ -1 +\overline{\sigma}_2\ &\ 1/ (\mathcal{D}_1 \lambda_2^2) & 0 \\
 -1\ &\ 0\ &\ 0 & 0 \\
  \rho v^2 - \hat{\chi}\
          &\ 0\ &\ 0 & -1 \\
  0\ &\   \rho v^2
           - \hat{\nu}\
             &\    -1 +\overline{\sigma}_2\ & 0
          \end{bmatrix},
\end{equation}
where
\be
 \hat{\chi} =
   \mathcal{D}_1(c_\theta^2 \lambda_1^2 + s_\theta^2 \lambda_3^2  + 3 \lambda_2^2 
    - 2\overline{\sigma}_2)  ,
\quad
 \hat{\nu} =
   \mathcal{D}_1[c_\theta^2 \lambda_1^2 + s_\theta^2 \lambda_3^2
   + \lambda_2^2(1 - \overline{\sigma}_2)^2],
\en
and $\overline{\sigma}_2 = \sigma_2 / (\mathcal{D}_1 \lambda_2^2)$ 
is a non-dimensional measure of the pre-stress.
The associated propagation condition is
\begin{equation}
 (q^2 + 1)[ \lambda_2^2 q^2 
   + (c_\theta^2 \lambda_1^2 + s_\theta^2\lambda_3^2 ) 
      - \rho v^2/\mathcal{D}_1] = 0.
\end{equation}

In other words, the situation is formally the same as that for
principal waves in Mooney-Rivlin material, see Section \ref{section_3_3}.
The conclusion is that the speed is given by 
\be 
\rho v^2 = \mathcal{D}_1(c_\theta^2 \lambda_1^2 + s_\theta^2\lambda_3^2  - \lambda_{2}^2 \eta^2),
\en
where $\eta$ is the real root of \eqref{secular_MR}.
\citet{Flav63} established this result at $\sigma_2 = 0$.
As he also showed, the situation gets more complicated for Mooney-Rivlin 
solids, because the wave is no longer plane polarized for a triaxial pre-stretch.

\CCLsubsubsection{Mooney-Rivlin solid.}

The Mooney-Rivlin 
strain energy density \eqref{mooney} at $\mathcal{D}_2 \ne 0$ does not lead to a decoupling 
of the saggital motion from the anti-saggital motion. 
However, as in Sections \ref{section_3} and \ref{section_4}, we find that the propagation condition factorizes,
here as
 \begin{equation} \label{biquadratic_mooney}
(q^2+1)(q^4 - Sq^2 + P) = 0,
\end{equation}
where
\begin{align}
& S = \left(\frac{1}{\gamma_{21}} + \frac{1}{\gamma_{23}}\right)X
       -\left(\frac{\gamma_{12}}{\gamma_{21}}
            + \frac{\gamma_{13}}{\gamma_{23}}\right)c_\theta^2
       -\left(\frac{\gamma_{31}}{\gamma_{21}}
            + \frac{\gamma_{32}}{\gamma_{23}}\right)s_\theta^2,
\nonumber \\
& P = (X - \gamma_{12}c_\theta^2 - \gamma_{32}s_\theta^2)
       (X - \gamma_{13}c_\theta^2 - \gamma_{31}s_\theta^2)
                                       /(\gamma_{21}\gamma_{23}).
\end{align}

\citet{Flav63} was the first to notice that $q_1 = \ii$ is an 
attenuation factor for non-principal waves in Mooney-Rivlin solids. 
\citet{Pich01} showed that a necessary and sufficient condition 
for the factorization \eqref{biquadratic_mooney} to occur is that the relations 
\eqref{relations_mooney} hold;
he also showed, completing earlier work by \citet{Will73}, 
that another factorization of the general bicubic \eqref{bicubic}
also occurs, this time for \emph{any strain energy function}, when two 
of the principal stretches of pre-strain are equal (equi-biaxial pre-strain). 
Finally note that $(q^2 + 1)$ always comes out as a factor in the propagation 
condition for inhomogeneous waves in Mooney-Rivlin solids, whatever 
the direction of propagation is (not necessarily in a principal
plane as here), see \citet{Dest02} for details.

Thanks to the factorization \eqref{biquadratic_mooney}, 
we can actually compute explicitly the  quantities 
$\omega_\text{I}$, $\omega_\text{II}$, $\omega_\text{III}$ 
of \eqref{omegas}.
Indeed, we now have
\be
q_1 = \ii, \qquad  q_2 q_3 = -\sqrt{P}, 
\qquad 
q_2 + q_3 = \ii \sqrt{2\sqrt{P}-S},
\en
so that
\begin{equation}
\omega_\text{I} = -\ii (1+\sqrt{2\sqrt{P}-S}),
\qquad
-\omega_\text{II} = \sqrt{P} + \sqrt{2\sqrt{P}-S},
\qquad
\omega_\text{III} = \ii \sqrt{P},
\end{equation}
leading to an explicit and exact form of the secular equation 
\eqref{secularImplicit},
\begin{equation} \label{explicitSecular}
 n\Bigl(1 + \sqrt{2\sqrt{P}-S}\Bigr)
   + \sqrt{P}\Bigl(m + \sqrt{P} +\sqrt{2\sqrt{P}-S}\Bigr)
    = 0.
\end{equation}

This result allows us to investigate the influence of pre-stress on 
surface wave propagation (see \citet{DOPR05} for an example)
and to address an important question in the study of surface stability:
how much can a Mooney-Rivlin half-space be compressed before it buckles?
In Section \ref{section_3} we found the critical stretch in a principal direction ($x_1$),
indicating the appearance of wrinkles parallel to $x_3$, 
see Figure \ref{figure_2} for a visualization. 
However, could it be that wrinkles appeared earlier in the compression,
in \emph{another} direction?
To answer this we take $X (=\rho v^2) = 0$ (onset of instability) and solve 
\eqref{explicitSecular} for $\lambda$, for each value of $\theta$. 

\begin{figure}  [htbp]
\centering 
 \epsfig{figure=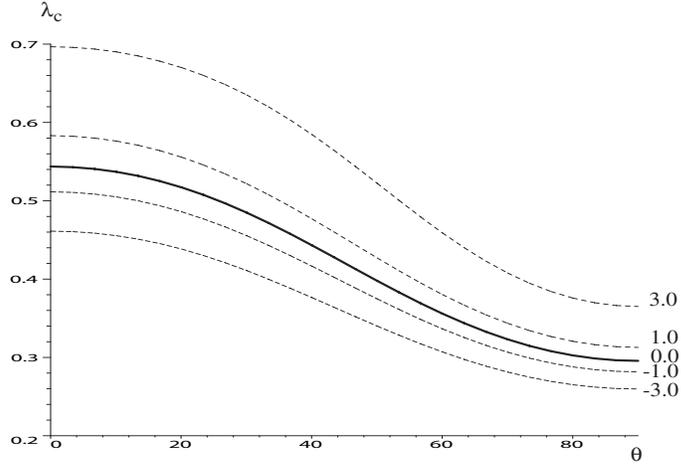, width=.66\textwidth, height=.46\textwidth}
 \caption{
 Critical stretch ratio of compression for a Mooney-Rivlin material as a 
 function of the angle between the normal to the wrinkles and the principal 
 axis of greatest compression. The Mooney-Rivlin solid is subject to a 
 finite plane strain compression ($\lambda_3 = 1$); 
 its material parameters are $\mathcal{D}_1 =2.0\mu$, $\mathcal{D}_2 = 0.8\mu$;
 the pre-stress normal to the boundary is given by  
 $\sigma_2 /\mu =$ -3.0, -1.0, 0.0, 1.0, 3.0
 ($\mu$ has the dimension of a stiffness).
  \label{fig_mr_non_principal}}
\end{figure}

For instance, take the case of the following plane pre-strain,
\be \label{plane_prestrain}
\lambda_1 = \lambda, \qquad \lambda_2 = \lambda^{-1}, 
\qquad \lambda_3 = 1,
\en
imposed on a half-space made of the Mooney-Rivlin solid with material 
parameters
\be \label{MR_example}
\mathcal{D}_1 =2.0\mu, \qquad \mathcal{D}_2 = 0.8 \mu,
\en
where $\mu$ has the dimension of a stiffness (the shear modulus 
of this Mooney-Rivlin solid is 
$(\mathcal{D}_1 + \mathcal{D}_2)/2 = 1.4 \mu$).
Figure \ref{fig_mr_non_principal} displays the values of the critical 
stretch ratio, measured in the principal direction $x_1$, 
for each angle $\theta$ and for several values of the pre-stress $\sigma_2$. 
The solid line corresponds to $\sigma_2=0$. 
At $\theta = 0$ we find the critical compression ratio $\lambda_\text{c}^0$ (say)
for wrinkles aligned along $x_3$;
hence the solid curve starts at  $\lambda_\text{c}^0 = 0.544$ as
expected from solving \eqref{principal_mr} with $v=0$ and 
$\lambda_1 = \lambda_\text{c}^0 = \lambda_2^{-1}$.
At $\theta \ne 0$ we find that the critical compression stretch is below
$\lambda_\text{c}^0$;
it follows that $\lambda_\text{c}^0$ is the absolute critical stretch 
of compression for our example \eqref{plane_prestrain}, \eqref{MR_example}.
Of course, for Mooney-Rivlin solids other than \eqref{MR_example}, or for 
pre-strains other than \eqref{plane_prestrain}, or for solids other than 
Mooney-Rivlin solids, we might end up with a different behaviour in compression. 
However the same analysis can be brought in each case to its conclusion,
with no additional difficulty. 

\CCLsubsubsection{Gent solid.}

We find that the elastic moduli \eqref{gammaBeta} for Gent materials
\eqref{Gent} are given in general by
\begin{align}
& \gamma_{i j} = C \dfrac{J_m}
  {J_m + 3 - \lambda_1^2 - \lambda_2^2 - \lambda_3^2}
     \lambda_i^2,
\nonumber \\
& 2 \beta_{i j} = C \dfrac{J_m}
  {J_m + 3 - \lambda_1^2 - \lambda_2^2 - \lambda_3^2}
     \left[ \lambda_i^2 + \lambda_j^2
 +  \dfrac{2(\lambda_i^2 - \lambda_j^2)^2}
  {J_m + 3 - \lambda_1^2 - \lambda_2^2 - \lambda_3^2}
\right].
\end{align}
Now consider that a half-space made of a Gent material with $J_m = 9.0$ 
is subject to a large
shear such that the boundary is the plane of shear 
(in Section \ref{section_4}, the boundary was the glide plane)
and the boundary $x_2=0$ is free of tractions. 
Then the  principal stretches are
\begin{equation}
  \lambda_{1} = \sqrt{1 + K^2/4} + K/2,  \qquad
\lambda_2 = 1, \qquad
  \lambda_{3} = \sqrt{1 + K^2/4} - K/2,
\end{equation}
and of course, $\lambda_1^2 + \lambda_2^2 + \lambda_3^2 -3 = K^2$.

\begin{figure}  [htbp]
\centering 
 \epsfig{figure=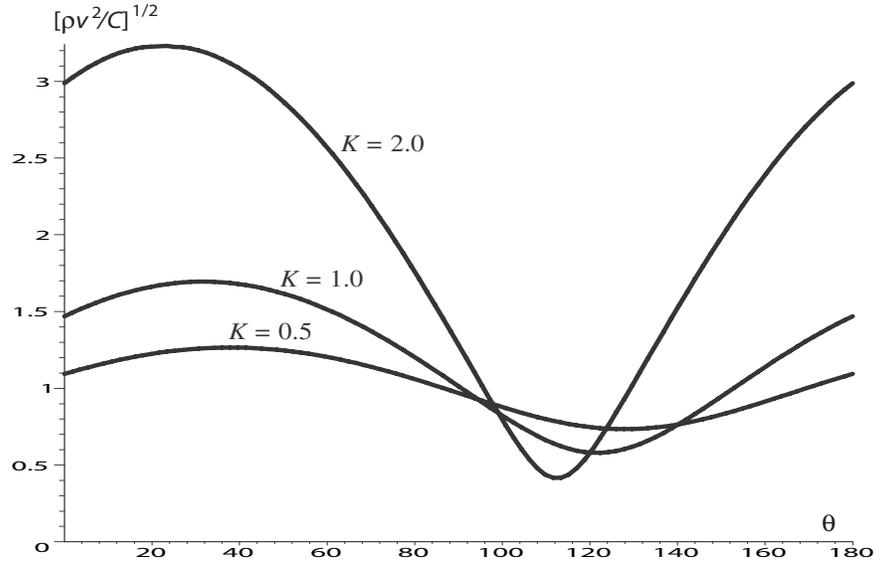, width=.85\textwidth, height=.55\textwidth}
 \caption{
 Wave propagating in the plane of shear of a semi-infinite sheared Gent solid
 with $J_m = 9.0$: 
 scaled speed $\sqrt{\rho v^2/C}$ as a function of the 
 angle between the 
 propagation direction and the direction of shear, for several values of the 
 amount of shear.
  \label{fig_gent_non_principal} }
\end{figure}

Figure \ref{fig_gent_non_principal} shows the variations of the surface wave speed in the plane of shear, 
with respect to the angle between the direction of propagation ($\hat{x}_1$) 
and the direction of shear ($X_1$). 
For a shear of amount $K=0.5$,
the principal axis of greatest stretch $x_1$ makes an angle $37.9^\circ$ with the 
direction of shear, and the principal axis of smallest stretch $x_3$ 
makes an angle $127.9^\circ$ with the direction of shear.
For $K=1$, those two angles are $31.7^\circ$ and $121.7^\circ$ , respectively;
for $K=2$, those two angles are $22.5^\circ$ and $112.5^\circ$ , respectively.
Clearly, the surface wave reaches extremal values along those directions, indicating 
that the principal directions can be determined acoustically.

    \CCLsection{Concluding remarks}

  This Chapter has presented several situations where it is possible to derive explicit 
secular equations in exact form and in polynomial form, for incremental waves 
propagating along the plane interface of one (or two) deformed semi-infinite 
hyperelastic solid (rigidly bonded solids). 
The subject of interface waves in general is broad and includes 
many other situations and geometries such as, to name but a few:
the addition of a layer of finite thickness (Love waves, Lamb waves, etc.), 
or of a fluid (Scholte waves for inviscid fluids, Stoneley waves for viscous fluids, etc.),
the consideration of an anisotropy due to families of extensible fibres, 
of cylindrical or spherical coordinates, 
or of curved boundaries, and so on.
Reading the works by \citet{Ogde34} and by \citet{Guz02} and the references therein 
gives a good overview of the vast panorama spanned by these other types of interface 
wave problems. 

This concluding Section expands on the reasons put forth to explain why it 
can be advantageous at times to seek explicit secular equations rather than to turn 
to numerics outright. 
Among other things, explicit secular equations in polynomial form
\begin{itemize}
	\item[\emph{(i)}] are easy to solve numerically, with the greatest precision required;
	\item[\emph{(ii)}] are sometimes surprisingly simple and short, in spite of
	a complicated or even unsolvable propagation condition;
	\item[\emph{(iii)}] lend themselves to simple asymptotic treatments, leading to approximate 
	analytical expressions for the wave speed;
	\item[\emph{(iv)}] account for all the solutions satisfying the propagation condition and the 
	boundary condition at the interface.
\end{itemize}
  
    \CCLsubsection{Numerics}
    \label{Section_6_1}

Point \emph{(i)} goes without saying because the numerical methods used to 
determine the roots of a polynomial are safe and robust. 
Of course, at most only one root corresponds to the actual solution 
and all the other roots must be discarded as being \emph{spurious}.
So, once a likely root is found numerically (that is a root $\rho v^2$ which 
is real and positive), it must be checked that at that speed the exact 
secular equation is satisfied. 
This routine check is simple enough to perform (solve the propagation condition,
find the corresponding impedance matrix, check that \eqref{secularM}
(for Rayleigh waves) or that \eqref{stoneley} (for Stoneley
waves) is satisfied).

Other numerical techniques to find the interface wave speed invoke the 
Barnett-Lothe-Stroh integral formalism (see \citet{Ting96} for an account),
computational algorithms for eigenvalue problems \cite{Tayl81},
or an algebraic Ricatti equation for the impedance matrix (see 
the Chapter by Fu in this book and the references therein) to 
determine $M(v)$ for any numerical value of $v$;
then $v$ is varied in order to satisfy the exact secular equation 
up to the desired precision.
This however is not always an easy task, as seen in the following example.

Consider the Mooney-Rivlin solid with material parameters 
\eqref{MR_example}, maintained in a state of static pre-strain, with 
\citep{RoSa99},
\be \label{mr_deformed_example}
\lambda_1 = \sqrt{3.695}, \qquad
\lambda_2 = \sqrt{0.7}, \qquad
\lambda_3 = (\lambda_1 \lambda_2)^{-1}, \qquad
\sigma_2 = 0.8 \mu.
\en
Take the interface between the solid and the vacuum to be the principal 
plane $x_2=0$ and study the propagation of a Rayleigh surface wave in a
direction close to $x_3$ ($\theta$ is close to $90^\circ$ in Figure 
\ref{fig_non_principal}.)

\begin{figure}  [htbp]
\centering 
 \epsfig{figure=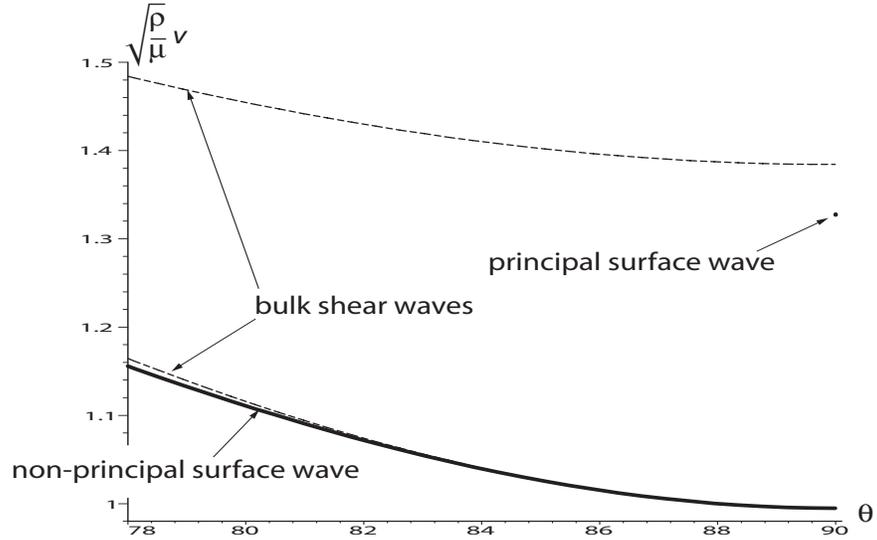, width=.85\textwidth, height=.55\textwidth}
 \caption{
 Wave propagation in the $x_2$ principal plane of semi-infinite Mooney-Rivlin solid \eqref{MR_example}
 deformed by \eqref{mr_deformed_example}:
 variations of scaled speed $\sqrt{\rho v^2/\mu}$ as a function of the angle, 
 near the $x_3$ direction.
  \label{fig_non_principal_mr_zoom} }
\end{figure}

Along $x_3$ we have a principal wave, travelling with a speed $v(90)$ say,
found from Section \ref{section_3_2}. 
Here \citep{DOPR05} we find that $v (90) \simeq 1.327 \sqrt{\mu /\rho}$.
On the other hand, a shear (homogeneous) bulk wave linearly polarized along $x_1$
travels with speed $v_1$ say, given by $\rho v_1^2 = \gamma_{31}$,
and a shear (homogeneous) bulk wave linearly polarized along $x_2$
travels with speed $v_2$ say, given by $\rho v_2^2 = \gamma_{32}$.
Here we find that $v_1 \simeq 1.384 \sqrt{\mu /\rho}$ and that 
$v_2 \simeq 0.995\sqrt{\mu /\rho}$, showing that the surface wave travels with a
speed which is \emph{intermediate} between those of the shear bulk waves. 
That principal wave is \emph{two-partial} and polarized in the ($x_2 x_3$) plane,
which is why it can afford to be faster than the shear bulk wave polarized in the 
$x_1$ direction.
Also, it is \emph{isolated}, in the sense that the transition toward a 
surface wave propagating in a direction $\theta \ne 90^\circ$ is abrupt,
because this latter wave is \emph{tri-partial} and must therefore travel 
with a speed $v(\theta)$ say, which is less than the speed of any 
homogeneous bulk shear wave, in particular less than $\sqrt{c_4/\rho}$
and less than $\sqrt{c_5 / \rho}$, 
where $c_4$ and $c_5$ are defined in \eqref{c_4}.
Figure \ref{fig_non_principal_mr_zoom} shows the variations of these 
3 speeds in the neighbourhood of the $x_3$ axis and makes those features clear. 

Now if our numerical method for the surface wave speed relies on 
using the known principal wave speed as an ``initial 
guess'', then it might run into difficulties here because of the 
isolation of the principal wave. 
This special feature is characteristic of strongly anisotropic crystals 
in linear elasticity. 
In incremental dynamics, it can occur at will, simply by deforming 
the solid sufficiently to create a strong strain-induced anisotropy.

Also, as is clear from the Figure, the surface wave travels with a speed which 
is extremely close to that of the bulk shear wave $\sqrt{c_4/\rho}$. 
Hence at $\theta = 89.99^\circ$, the speed of the former is given by
\be \label{example_8999}
\sqrt{\rho v^2 / \mu} = 0.9948641879596778\ldots,
\en
whilst the speed of the latter is given by
\be
\sqrt{c_4 / \mu} =      0.9948641879596860\ldots.
\en
Consequently, if a numerical scheme for the surface wave speed 
relies on increasing $v$ in small steps until $\text{det }M(v) = 0$ 
is satisfied to any desired precision, then it will have to be pushed 
to the 14th significant digit to insure that the speed does not 
correspond to a bulk wave! 
Here the explicit polynomial secular gives only two positive roots, and it 
is a routine check to verify that only \eqref{example_8999} satisfies the exact secular equation.
  
    \CCLsubsection{Polynomials}

For an interface wave propagating \emph{in any direction in a deformed solid}, the propagation condition 
is a sextic in general. 
Such is for instance the case for a wave propagating in any direction in the glide plane 
of the sheared block in Figure \ref{figure_4}.
Although a sextic is unsolvable analytically, it is nonetheless possible to find a 
polynomial secular equation for a Rayleigh surface wave, see \citet{Tazi89} or \citet{Ting04} for details.
The polynomial turns out to be of degree 27 in $\rho v^2$. 
For Stoneley interface waves it does not seem practical to look for a polynomial 
secular equation. 

For an interface wave propagating \emph{in any direction in a principal plane} of 
pre-de\-for\-ma\-tion, the propagation condition 
is the bicubic \eqref{bicubic}. 
Although formulas exist for the roots of a cubic, the analytical resolution of \eqref{bicubic}
proves to be complicated  because it is not known a priori whether the roots are purely imaginary or 
have a non-zero real part and accordingly, which formula should be used. 
However, the polynomial explicit secular equation for a Rayleigh surface wave is of 
degree 12 in $\rho v^2$ as seen earlier. This is also the case in the symmetry plane of an 
orthorhombic or monoclinic crystal. 
In the symmetry plane of a cubic crystal, the degree of the polynomial comes down to 10, 
as noted by \citet{Tayl81}.
For a two-partial surface wave, coupled to electrical fields through piezoelectricity 
in 2mm crystals, the polynomial is also of degree 10 in $\rho v^2$ for metallized boundary conditions
(there the propagation condition is also a bicubic, see \cite{CoDe05}). 

For a two-partial (non-principal) interface wave \emph{polarized in a principal plane} of pre-deformation, 
the propagation condition is a quartic,
\begin{equation}
 q^4 + d_3 q^3 + d_2 q^2 + d_1 q + d_0 =0, 
 \end{equation}
say (see \eqref{quartic_shear} for the case of simple shear).
In contrast to the case of the bicubic above, we know here what the form of the 
roots should be, because the roots of a quartic are either two pairs of complex conjugate
numbers, or one double real root with one pair of complex conjugate numbers, or 
four real roots. 
Here only the first scenario is acceptable in order to satisfy the decay condition 
\eqref{decay} and thus a single formula, always valid, is required; 
it can be found in the textbooks. 
First introduce in turn the quantities $r$, $s$, and $h$, defined by
\be \label{rsh}
 r = d_2 - \textstyle{\frac{3}{8}} d_3^2,
\quad
 s = d_1 - \textstyle{\frac{1}{2}} d_2d_3
      + \textstyle{\frac{1}{8}} d_3^2,
\quad
 h = d_0 - \textstyle{\frac{1}{4}} d_1 d_3
       + \textstyle{\frac{1}{16}} d_2 d_3^2
        - \textstyle{\frac{3}{256}} d_3^4,
\en
the quantities $\lambda$ and $\phi \in ]0, \pi/3]$, defined by
\be \lambda =
  \textstyle{\frac{1}{27}} \left(12 h + r^2\right)^{3/2}, \qquad
\phi =\textstyle{\frac{1}{3}} \arccos
  \left[ \frac{1}{2 \lambda}
     \left(\textstyle{\frac{2}{27}} r^3
          + s^2 - \textstyle{\frac{8}{3}} r h\right)\right],
\en 
and the quantities $z_1$, $z_2$, and $z_3$, defined by
\begin{align}
& z_1 = 2 \lambda^{1/3} \cos (\phi)
         - \textstyle{\frac{2}{3}} r,
\nonumber \\
& z_2 = 2 \lambda^{1/3} \cos (\phi + 2 \pi/3)
        - \textstyle{\frac{2}{3}} r,
\nonumber \\
& z_3 = 2 \lambda^{1/3} \cos (\phi + 4 \pi/3)
         - \textstyle{\frac{2}{3}} r.
\end{align}
Then  the qualifying roots are
\begin{align} \label{formula_2}
& p_1 = \textstyle{\frac{1}{2}}\text{sign}(s)\sqrt{z_1}
          - \textstyle{\frac{1}{4}} d_3   
           + \textstyle{\frac{1}{2}} \ii (\sqrt{-z_2} + \sqrt{-z_3}),
\notag \\
& p_2 = -\textstyle{\frac{1}{2}}\text{sign}(s)\sqrt{z_1}
          - \textstyle{\frac{1}{4}} d_3   
           + \textstyle{\frac{1}{2}} \ii (\sqrt{-z_2} - \sqrt{-z_3}),
\end{align}
where  $\text{sign}(s)$ equals 1 if $s$ is non-negative and $-1$
otherwise.
These formulas are perfectly well handled by a computer algebra 
system, so that $M$, and ultimately the exact secular equation, can be found. 
It then means that any interface wave problem can be solved exactly.
Nonetheless it might still be rewarding to look for the polynomial secular equation,
to check whether they turn out to be simple. 
For instance, the polynomial secular equation for Rayleigh waves is a \emph{quartic in the 
squared speed} when the solid is sheared (Section \ref{section_4_2_1}), or 
tri-axially stretched and then sheared \citep{DeOg05}, or when the 
wave is polarized in the symmetry plane of a crystal \citep{Curr79}.
Now consider a flexural wave travelling along the free edge of a 
thin (Love-Kirchhoff) orthotropic plate where the edge makes an arbitrary 
angle with a principal axis of symmetry. 
\citet{ThAN02} show that the corresponding propagation condition is a 
quartic, but turn to a numerical resolution. 
However \citet{Fu03} shows that the polynomial secular equation is just a 
\emph{cubic in the squared speed}.
Finally consider the case of a piezoacoustic Bleustein-Gulyaev surface wave 
in a rotated $Y$-cut about the $Z$ axis for $\overline{4}$ crystals. 
There also the propagation condition is a quartic, but the fundamental equations
reveal that the polynomial secular equation for metallized boundary conditions is simply 
a \emph{quadratic in the squared speed} \citep{CoDe04}.
Clearly it is a worthy enterprise to unearth these polynomials rather than use numerics
or the formulas \eqref{rsh}-\eqref{formula_2}.

    \CCLsubsection{Approximate expressions}

Once the polynomial secular equation is established, in the form
$\mathcal{P} (v) = 0$, say, it is a straightforward matter to derive an analytical
approximation for $v$. 
Calling $v_0$ an initial approximation, we find in the first order that
\be \label{approximation}
v \simeq v_0 - \mathcal{P}(v_0)/\mathcal{P}'(v_0).
\en
Of course we must choose $v_0$ judiciously for an optimal expression. 

\begin{figure} [htbp]
 \centering
  \mbox{\epsfig{figure=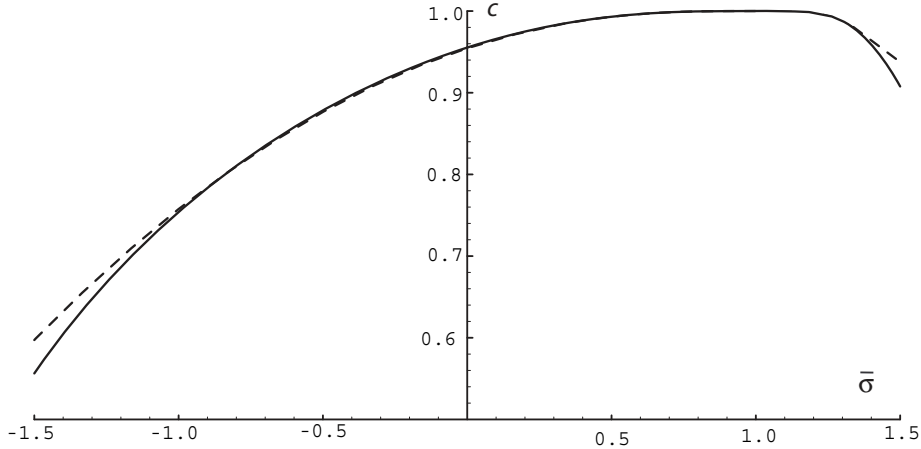, width=.9\textwidth}}
\caption{{\small Surface wave speed in an incompressible 
solid subject to hydrostatic pressure only. 
Solid plot: exact calculation, dashed plot:
approximate calculation.} \label{fig_approx}}
\end{figure}

Here we work out a simple example. 
We take a solid half-space subject to a \emph{hydrostatic pressure} only,
so that $\lambda_1 = \lambda_2 = \lambda_3 \equiv \lambda$ say, 
and $\sigma_1 = \sigma_2 = \sigma_3 \equiv \sigma$ say.
From the incompressibility constraint \eqref{incompressibility}, 
$\lambda = 1$ follows and we  have a \emph{pre-stressed},
but \emph{unstrained}, solid.
It is thus \emph{isotropic} and a surface wave propagates with the same speed in 
every direction. 
To derive the secular equation, we specialize for instance \eqref{secular_principal}
to the isotropic case. 
We find that $\gamma_{12} = \gamma_{21} = \beta_{12} = \mu_0$,
the infinitesimal shear modulus, and the exact secular equation is 
\citep{DoOg90}
\be \label{exact_hydro}
f(\eta) = \eta^3 + \eta^2 + (3 - 2\overline{\sigma} )\eta 
 - (1-\overline{\sigma})^2 = 0,
 \en
where $\overline{\sigma} \equiv \sigma / \mu_0$ and 
$\eta = \sqrt{1 - \rho v^2 / \mu_0}$.
Calling $c \equiv \sqrt{\rho v^2 / \mu_0}$ a dimensionless measure 
of the wave speed, and multiplying \eqref{exact_hydro} by $f(-\eta)$,
gives the polynomial secular equation \citep{DoOg90},
\be
\mathcal{P} (v) \equiv 
 c^6 - 4(2 - \overline{\sigma})c^4 
   + 6(\overline{\sigma} -2)^2 c^2
     + (\overline{\sigma}^2 - 4)(\overline{\sigma} - 2)^2 = 0.
\en
In the region where $c$ is close to 1, the first order approximation 
\eqref{approximation} gives $c \simeq 1 - \mathcal{P}(1)/\mathcal{P}'(1)$,
that is
\be
c \simeq \dfrac{21 - 28\overline{\sigma}
      + 6\overline{\sigma}^2 + 4\overline{\sigma}^3 
         - \overline{\sigma}^4}
         {22 - 32\overline{\sigma} + 12\overline{\sigma}^2}.
\en
In particular we find that for an unstressed incompressible solid, 
$c \simeq 21/22 = 0.9545\ldots$, which is a good approximation to the 
value found from the exact equation 0.9553\ldots (Lord Rayleigh, 1885).
On Figure \ref{fig_approx} we superpose the approximate and exact curves,
and find good agreement in the $\overline{\sigma} = -1$ to 
$\overline{\sigma} = 1$ region.
Beyond that range, $c$ departs from the neighbourhood of 1, and a better
choice must be made for the initial guess.
This point is not developed further here but it is clearly exposed in the 
paper by \citet{Mozh91}.

    \CCLsubsection{Other waves}
 
\citet{Tayl81} and \citet{Tazi89} both remarked that many of the roots 
to an explicit secular equation in polynomial form have a physical origin, 
because they correspond to a motion satisfying both the equations of motion 
and the boundary conditions at the interface. 
In particular for the solid/vacuum interface, the polynomial 
provides the parameters not only for the Rayleigh surface wave, but also 
for pseudo-surface waves, for Brewster reflection of bulk waves, for the reflection of inhomogeneous waves, 
for ``organ-pipe'' modes, etc.

One of the most important root is that corresponding to a \emph{leaky 
surface wave} (if it exists), whose energy is diffused slowly into the half-space, 
and for which the classical methods evoked in Section \ref{Section_6_1}
run into major difficulties. 

\begin{figure} [htbp]
 \centering
  \mbox{\epsfig{figure=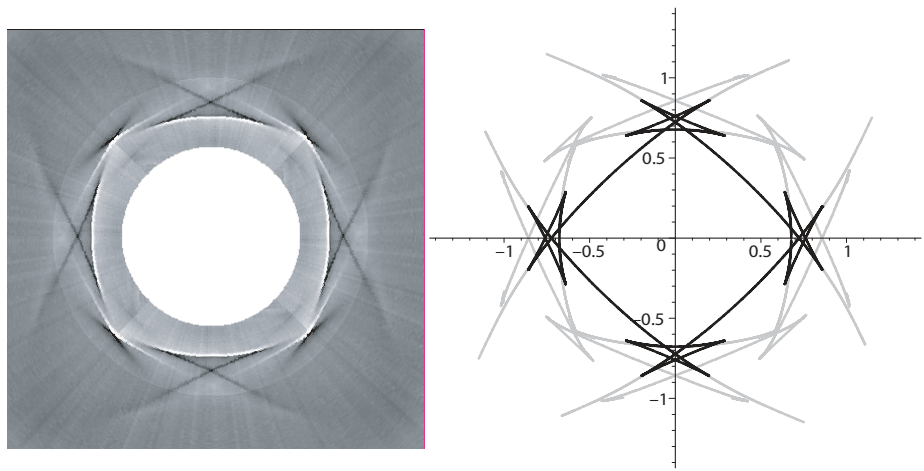, width=.9\textwidth}}
\caption{{\small Wavefronts on the surface of copper cut along a symmetry plane. 
Left: experimental results obtained by LASER impact  \citep{Huet06}. 
Right: predictions derived from the secular equation in polynomial form,
showing in black, the Rayleigh wave wavefront and in grey, the wavefronts
due to leaky waves and interference of inhomogeneous waves.} \label{fig_huet}}
\end{figure}

Another application for this wealth of information is that many more \emph{wavefronts}
can be found, in particular those observed in the neighbourhood of cusps, which 
correspond to the interference of inhomogeneous plane waves \citep{Huet06}.
Figure \ref{fig_huet} shows the remarkable correspondence obtained between experiment 
and predictions, for wavefronts propagating on the surface of a copper sample (cubic 
symmetry). 
Here the explicit secular equation proves useful because it is easy to 
differentiate, as required for the derivation of the curve.

    \bibliographystyle{plainnat}
    \bibliography{Contribs_destrade}

\end{document}